\documentclass[twocolumn,superscriptaddress,floatfix]{revtex4-2}
\usepackage{graphicx}
\usepackage{dcolumn}
\usepackage[utf8]{inputenc}
\usepackage[T1]{fontenc}
\usepackage{mathptmx}
\usepackage{etoolbox} 
\usepackage{gnuplottex}
\usepackage{mathtools}
\usepackage{float}
\usepackage{array}
\usepackage[english]{babel}
\usepackage[utf8]{inputenc}
\usepackage[table,dvipsnames]{xcolor}
\usepackage{amsfonts}  
\usepackage{graphicx}
\graphicspath{ {./Figures/} }
\usepackage{dsfont}
\usepackage[colorlinks=true, linkcolor=blue, citecolor=blue]{hyperref}
\usepackage{algorithm}
\usepackage[noend]{algpseudocode} 
\usepackage{mathtools} 
\usepackage{bm} 
\usepackage{xspace}
\usepackage{listings}
\usepackage{tikz}
\usepackage{bbding}
\usepackage{physics}
\usepackage{placeins} 
\usepackage{standalone}
\usetikzlibrary{arrows.meta}
\usetikzlibrary{shapes}

\newcommand{\be}{\begin{equation}}
\newcommand{\ee}{\end{equation}} 

\makeatletter
\def\@email#1#2{%
 \endgroup
 \patchcmd{\titleblock@produce}
  {\frontmatter@RRAPformat}
  {\frontmatter@RRAPformat{\produce@RRAP{*#1\href{mailto:#2}{#2}}}\frontmatter@RRAPformat}
  {}{}
}%
\makeatother

\begin{document}





\title[]{  
\emph{Ab Initio} Polaritonic Chemistry on Diverse Quantum Computing Platforms: \\
 Qubit, Qudit, and Hybrid Qubit-Qumode Architectures
}

\author{Even Chiari}
\affiliation{Laboratoire de Chimie Quantique de Strasbourg, Institut de Chimie,
CNRS/Université de Strasbourg, 4 rue Blaise Pascal, 67000 Strasbourg, France} 
\author{Wafa Makhlouf}
\affiliation{Laboratoire de Chimie Quantique de Strasbourg, Institut de Chimie,
CNRS/Université de Strasbourg, 4 rue Blaise Pascal, 67000 Strasbourg, France} 
\author{Lucie Pepe}
\affiliation{Laboratoire de Chimie Quantique de Strasbourg, Institut de Chimie,
CNRS/Université de Strasbourg, 4 rue Blaise Pascal, 67000 Strasbourg, France} 
\author{Emiel Koridon} 
\affiliation{Theoretical Chemistry, Vrije Universiteit, De Boelelaan 1083, NL-1081 HV, Amsterdam, The Netherlands}
\affiliation{Instituut-Lorentz, Universiteit Leiden, P.O. Box 9506, 2300 RA Leiden, The Netherlands}
\author{Johanna~Klein}
\affiliation{ICGM, Univ Montpellier, CNRS, ENSCM, Montpellier, France} 
\author{Bruno Senjean} 
\affiliation{ICGM, Univ Montpellier, CNRS, ENSCM, Montpellier, France}
\author{Benjamin Lasorne} 
\affiliation{ICGM, Univ Montpellier, CNRS, ENSCM, Montpellier, France}
\author{Saad Yalouz}
\thanks{corresponding author: \href{mailto:saad.yalouz@cnrs.fr}{saad.yalouz@cnrs.fr}} 
\affiliation{Laboratoire de Chimie Quantique de Strasbourg, Institut de Chimie,
CNRS/Université de Strasbourg, 4 rue Blaise Pascal, 67000 Strasbourg, France} 

\date{\today}

\begin{abstract}

In the field of \textit{ab initio} polaritonic chemistry, the use of quantum computers may offer, in the long term, a promising computational pathway for simulating and exploring novel cavity-induced light-matter phenomena. 
Trying to export such a problem onto emerging quantum computers, however, raises fundamental questions. \linebreak
A central one is how to efficiently represent both fermionic and bosonic degrees of freedom on the same platform, in order to develop computational strategies that can accurately capture strong electron-photon correlations at a reasonable cost for implementation on near-term hardware.
Given the hybrid fermion-boson nature of polaritonic problem, one may legitimately ask then: should we rely exclusively on conventional qubit-based platforms, or consider alternative computational paradigms better suited to the high dimensionality of light-matter systems?
To explore this, we investigate in this work three strategies for representing and solving cavity quantum electrodynamics problems, namely: qubit-based, qudit-based, and hybrid qubit-qumode approaches.
For each platform, we design compact, physically motivated quantum circuit ans\"{a}tze and integrate them within the state-averaged variational quantum eigensolver to compute multiple polaritonic eigenstates simultaneously.
A key element of our approach is the development of compact electron-photon entangling circuits, tailored to the native capabilities and limitations of each hardware architecture.
We benchmark all three strategies on a cavity-embedded H\textsubscript{2} molecule, resolving the three lowest-energy polaritonic states and reproducing characteristic phenomena such as light-induced avoided crossings and conical intersections.
Our results show that each platform achieves comparable accuracy in predicting polaritonic eigen-energies and eigenstates.
However, with respect to quantum resources required (\textit{i.e.} number of entangling gates  and quantum information units), the hybrid qubit-qumode approach offers the most favorable tradeoff between resource efficiency and accuracy, followed closely by the qudit-based method.
Both of which outperform the conventional qubit-based strategy.
\linebreak Our work presents a hardware-conscious comparison of quantum encoding strategies for polaritonic systems and highlights the potential of higher-dimensional quantum platforms to simulate complex light-matter systems.

\end{abstract}

\maketitle

\section{Introduction}
 
Polaritonic chemistry investigates the emergent behaviors that arise when molecular systems or {extended} materials interact with confined, quantized electromagnetic fields within optical cavities. This field, primarily driven by pioneering experiments from Ebbesen’s group~\cite{ebbesen2016hybrid,ebbesen2023introduction}, has garnered significant attention in both the chemistry and physics communities due to the wide array of exotic effects it can generate. 
Indeed, numerous experimental studies have demonstrated that cavity-confined light-matter interactions can influence a variety of molecular and material processes, including chemical reactivity~\cite{hutchison2012modifying,thomas2016ground,lather2019cavity}, charge transfer~\cite{sandik2024cavity,schafer2019modification}, and energy transport~\cite{zhong2016non,rozenman2018long}, among others.
Interestingly, despite significant experimental progress, theoretical modeling of polaritonic chemistry still represents a major challenge. 
This difficulty is closely tied to the complexity of the underlying quantum phenomena that occur between light and matter within a cavity.

Polaritonic effects emerge from the formation of hybrid light-matter quantum states, which exhibit properties distinct from those of their individual atomic/photonic components.
Predicting the properties of these states requires a quantum many-body description accounting for the interaction between the electronic structure of molecules (fermions) and the quantized photonic field within the cavity (bosons). 
Solving the corresponding many-body (electron+photon) Schrödinger equation is a genuine Quantum ElectroDynamics (QED) problem~\cite{cohen1987photons, cohen1996processus}, which presents substantial theoretical and computational challenges (regarding notably the dimensionality of the many-body Hilbert space involved).
To address these challenges, simplified models from quantum optics have been widely explored~\cite{shore1993jaynes, jaynes1963comparison, cummings1965stimulated}.
Although these models provide valuable insights into the underlying physics in the cavity (see, for example,~\cite{wellnitz2022disorder, wellnitz2021quantum, larson2021}), the latter often oversimplify the many-electron nature of the trapped molecules, leading to a poor description of their intrinsic chemical detail. \linebreak
To overcome these theoretical limitations, recent works have revisited the field of polaritonic chemistry by combining \textit{ab~initio} electronic structure methods with a rigorous quantum description of the photonic field. 
In this context, various methodological advancements have emerged,
including polaritonic ``QED-like version'' of many famous quantum chemistry methods.
These advancements include the development of the QED Hartree-Fock theory~\cite{haugland2020coupled,riso2022molecular}, coupled-cluster theory~\cite{haugland2020coupled, mordovina2020polaritonic}, density functional theory~\cite{ruggenthaler2014quantum, tokatly2013time}, perturbative approaches~\cite{bauer2023perturbation, cui2024variational, moutaoukal2025strong}, configuration-interaction methods~\cite{mctague2022non,vu2024cavity}, and density matrix renormalization group techniques~\cite{matousek2024polaritonic}, to cite a few examples (see Refs.~\cite{bhuyan2023rise, mandal2023theoretical, ruggenthaler2023understanding, foley2023ab} for extensive reviews).

All of these \textit{ab initio} polaritonic methods were originally developed to provide a more accurate description of strong electron-photon correlations through simulations executed on conventional \textit{``classical computers''}. 
More recently, and in parallel with these classical approaches, a growing number of studies have turned to \textit{quantum computers} to address the same challenges. 
Significant contributions in this direction include the development of quantum algorithms based on the anti-Hermitian contracted Schrödinger equation~\cite{warren2025quantum}, as well as the formulation of polaritonic Unitary Coupled Cluster theory~\cite{pavosevic2021polaritonic}, which has been implemented within the framework of the Variational Quantum Eigensolver (VQE) and applied to noisy quantum platforms~\cite{hassan2024simulating}. 
Along similar lines, it is also worth noting the existence other works that have explored general fermion-boson problems on quantum computers, such as those reported in Refs.~\cite{casanova2011quantum, mezzacapo2012digital, lamata2014efficient, macridin2018digital, kumar2023digital}.

 Remarkably, it is interesting to note that, despite the inherently quantum nature of \textit{ab initio} polaritonic chemistry, relatively few works have been devoted to the development of quantum algorithms in this area. 
 One possible explanation lies in the theoretical and practical challenges of representing and simulating hybrid electron-photon systems within the conventional \textit{qubit-based} computational framework.

Indeed, \textit{qubit-based} quantum computing has been the most dominant computational paradigm considered in the simulation of many-body systems, more specifically for the treatment of fermionic systems.
This success is largely attributed to the natural correspondence between the computational basis of qubits and the fermionic Fock space, making it a convenient and widely adopted method for electronic structure problems.
However, polaritonic chemistry inherently involves both fermionic (electrons) and bosonic (photons) degrees of freedom. 
While mapping bosonic modes onto qubits is theoretically feasible, encoding them with two-level systems demands substantially more quantum resources than fermions and introduces greater complexity in quantum circuits (see Refs.~\cite{somma2002simulating, veis2016quantum, peng2023quantum,knapik2025quantum}).
These observations naturally prompt several fundamental questions: \linebreak Is the conventional \textit{qubit-based} approach adequate for describing hybrid light-matter systems? 
Are there alternative computational strategies that could offer a better fit for electron-photon interactions while maintaining manageable quantum resource costs?

Motivated by these questions, the goal of this prospective work is to move beyond the conventional \textit{qubit-based} approach and explore a broader range of quantum computing platforms to treat the \textit{ab initio} polaritonic chemistry problem.\linebreak
Our ambition is to theoretically assess whether alternative approaches employing higher-dimensional quantum units of information (i.e., \textit{qudits} and \textit{qumodes}) may offer practical advantages for simulating such hybrid light-matter problems.
Building on recent developments in quantum algorithms, specifically those based on the State-Averaged Variational Quantum Eigensolver (SA-{VQE})~\cite{nakanishi2019subspace,yalouz2021state,yalouz2022analytical}, we aim to evaluate the potential of different quantum platforms to generate circuit ansatze capable of encoding the complex fermion-boson entanglement present in multiple polaritonic light-matter states (simultaneously). 
To this end, we will investigate three different platforms that utilize either \textit{i)}~qubits, \textit{ii)}~qudits, or \textit{iii)~}hybrid qubit-qumode units of information.

Indeed, one promising direction for overcoming the limitations of traditional \textit{qubit-based} systems is the use of \textit{qudits}, which represents a discrete variable whose number of quantum states extend beyond conventional binary frameworks (with dimension $d > 2$).
As we will show later on in this work, such a unit of information could define a very appealing way of encoding bosonic states in a more compact manner. \linebreak
Initially introduced by Ekert in 1991 for quantum cryptography applications \cite{Ekert1991qudits}, the concept of qudits has since been explored in several studies \cite{Howard2012qudits, Jamil_Daboul_2003qudits, Gao2020qudits, ermann2020qudits, Ivanov2012qudits}. 
In particular, Gotyal \cite{Goyal2014qudits} provided crucial validation for the practical use of qudits in quantum communication, opening new possibilities for processing high-dimensional quantum information. 
This approach has been shown to allow for more advanced manipulation of quantum states, which could enhance the performance of quantum algorithms \cite{Lu2020qudits, LUO2014qudits}.
The practical realization of qudits requires platforms capable of supporting and manipulating high-dimensional quantum states. 
Several promising physical systems have been identified for this purpose, including cold atoms trapped in optical lattices \cite{Dong2023qudits, Sawant_2020qudit, Lindon_2023qudits}, trapped ions \cite{Ringbauer_2022qudits, Low_2020qudits, Nikolaeva2024qudits}, and photonic systems \cite{Wang_2020qudits, Márton2024qudits}.
While promising, the development of qudit-based computing is still in its infancy, and ongoing active research is being carried out to fully understand the computational advantages of qudits compared to qubits~ (see for example Refs.~\cite{kiktenko2025colloquium,jankovic2024noisy}).

As another alternative, \textit{hybrid qubit-qumode} quantum computing architectures represent a particularly attractive route to describe polaritonic problems.
This novel computational framework is based on the integration of two types of quantum information units: \textit{discrete variables}, which are \textit{qubits}, that can interact with \textit{continuous variables}, known as \textit{“qumodes”}~\cite{casanova2011quantum, mezzacapo2012digital, lamata2014efficient, macridin2018digital}. 
Qumodes represent genuine quantum harmonic oscillators, with the advantage of naturally encoding bosonic degrees of freedom, as illustrated in various applications (see, for example, Refs.~\cite{kalajdzievski2018continuous, yalouz2021encoding}).
While ``purely'' qumode-based quantum computers already exist (primarily photonic platforms), recent experimental advancements suggest the possibility to develop hybrid hardware platforms that combine both qubit and qumode units of information, as demonstrated with superconducting circuits and trapped ions (see Refs.~\cite{liu2024hybrid, crane2024hybrid, araz2024toward, stavenger2022c2qa}).
In these systems, qubits (e.g., transmons, ion states) couple with quantum harmonic oscillators such as microwave cavities or nanomechanical resonators, which inherently carry physical bosonic excitations.
Situated at the intersection of digital and analog computation, these hybrid platforms~\cite{casanova2011quantum, mezzacapo2012digital, lamata2014efficient, macridin2018digital, kumar2023digital} present a promising approach for more versatile quantum computations, including the description of hybrid fermion-boson interactions. 

Motivated by these recent experimental and theoretical developments, in this work, we investigate how one could leverage \textit{qudits} and \textit{qubit-qumode} architectures to move beyond conventional \textit{qubit-based} approaches for applications in polaritonic chemistry.
To this end, we design a set of physically motivated quantum circuit ans\"{a}tze tailored to each platform and integrate them within the SA-VQE framework.
Our goal is to describe multiple low-energy polaritonic states within a single circuit, capturing characteristic phenomena such as light-induced avoided crossings and conical intersections.
Special attention is devoted to the encoding of bosonic degrees of freedom, as well as to the implementation of compact fermion-boson entangling operations adapted to each hardware platform.
These approaches are systematically benchmarked in terms of accuracy and resource efficiency on an \textit{ab initio} system, namely: H$_2$ embedded in an optical cavity as modeled by the Pauli–Fierz Hamiltonian~\cite{cohen1987photons, cohen1996processus}.

The paper is organized as follows.
In Sec.~\ref{sec:II}, we outline the theoretical background of \textit{ab initio} polaritonic chemistry, introduce the Pauli-Fierz Hamiltonian, and illustrate how strong light-matter coupling can induce non-trivial phenomena such as avoided crossings and conical intersections.
In Sec.~\ref{sec:III}, we present the quantum computing framework underlying our study. 
After reviewing the SA-VQE algorithm, we detail the encoding strategies for each quantum computing platform, focusing on the treatment of strongly correlated electron-photon states. 
Circuit design principles and quantum resource considerations are also discussed.
In Sec.~\ref{sec:IV}, we report our numerical simulations and compare the performance of the different approaches.
Finally, we summarize our findings and discuss future directions in Sec.~\ref{sec:V}.




\section{Ab INITIO Polaritonic Quantum Chemistry}
\label{sec:II}
In this first section, we introduce the theoretical background for describing \textit{ab initio} polaritonic chemistry, as formulated within the Pauli-Fierz Hamiltonian. 
Using the illustrative example of H$_2$ in a cavity, we then discuss how the formation of hybrid light-matter states gives rise to exotic physical phenomena, including light-induced avoided crossings and conical intersections. 
The simple (yet rich) H$_2$ system will serve as the primary testbed for the quantum computing developments explored in the subsequent sections.

\subsection{Pauli-Fierz Hamiltonian}

Building on the extensive body of work in \textit{ab initio} polaritonic chemistry, we adopt the widely used Pauli-Fierz Hamiltonian, which provides a non-relativistic \textit{ab initio} description of how the electronic structure of a molecule interacts with photons in a cavity. 
The derivation of this Hamiltonian is a complex process grounded in QED theory, and the detailed steps, starting from the fundamental QED Lagrangian, are thoroughly discussed in reference textbooks~\cite{cohen1987photons,cohen1996processus}. 
More recently, accessible reviews have outlined the key steps in transitioning from the minimal coupling ``p·A'' Hamiltonian
to the Pauli-Fierz form, which we adopt in this work (see reviews~\cite{mandal2023theoretical,foley2023ab}). 
We encourage interested readers to consult these references for a comprehensive discussion of the derivation process, while we focus here on the final form of the Pauli-Fierz Hamiltonian, which reads as follow in the Born-Oppenheimer approximation:
\begin{equation}\label{eq:Pauli_fierz_ham}
\begin{split}
    \hat{H} &=\hat{H}_{\mathrm{e}} 
    \\&+ \sum_\alpha \omega_{ \alpha} \hat{b}_\alpha^{\dagger} \hat{b}_\alpha
    \\&- \sum_\alpha \sqrt{\frac{\omega_\alpha}{2}} (\boldsymbol{\lambda}_\alpha \cdot \bold{\hat{d}} ) \left(\hat{b}_\alpha^{\dagger} 
    + \hat{b}_\alpha\right)
    + \frac{1}{2} \sum_\alpha ({\boldsymbol{\lambda}}_\alpha \cdot \bold{\hat{d}})^2.
\end{split}
\end{equation}

In this definition, the first term $\hat{H}_e$ corresponds to the regular electronic structure Hamiltonian from quantum chemistry defined in second quantization by
\begin{equation}
\hat{{H}}_e
=
\sum_{pq} h_{pq} \hat{E}_{pq} + \dfrac{1}{2}
\sum_{pqrs}
g_{pqrs}\hat{e}_{pqrs} + \mathcal{E}_{nuc},
\end{equation}
where $\mathcal{E}_{nuc}$ is the nuclei-nuclei repulsion, and $h_{pq}$ and $g_{pqrs}$ are the one- and two-electron integrals defined as 
\begin{align}
h_{p q} &=\int \phi_p^*\left(\mathbf{r}\right)\left(-\frac{1}{2}{\nabla}^2-\sum_{A} \frac{Z_A}{|\bold{r} - \bold{r}_{A}|}\right) \phi_q\left(\mathbf{r}\right) \mathrm{d} \mathbf{r}, \\
g_{p q r s} &=\iint \phi_p^* \left(\mathbf{r}_1\right) \phi_r^*\left(\mathbf{r}_2\right) \frac{1}{|\bold{r}_1 - \bold{r}_2|} \phi_q\left(\mathbf{r}_1\right) \phi_s\left(\mathbf{r}_2\right) \mathrm{d} \mathbf{r}_1 \mathrm{~d} \mathbf{r}_2,
\end{align} 
where $\lbrace \phi_p \rbrace$ represent the finite  basis set of spatial molecular orbitals.
The one- and two-electron spin-free excitation operators, $\hat{E}_{pq}$ and $\hat{e}_{pqrs}$, are defined such as~\cite{helgaker2013molecular}
\begin{equation} 
        \hat{E}_{pq}   = \sum_{\sigma} \hat{a}_{p\sigma}^\dagger \hat{a}_{q\sigma}, \quad \text{and} \quad \hat{e}_{pqrs}  = \sum_{\sigma,\tau} \hat{a}_{p\sigma}^\dagger \hat{a}_{r\tau}^\dagger \hat{a}_{s\tau}\hat{a}_{q\sigma}, 
\end{equation}
where $\hat{a}_{p\sigma}^\dagger/\hat{a}_{p\sigma}$ are electronic creation/annihilation operators in a spin-orbital with spatial part $\phi_p$ and spin state $\sigma \in \lbrace \uparrow,\downarrow \rbrace$.

In the Pauli-Fierz Hamiltonian definition given in Eq.~(\ref{eq:Pauli_fierz_ham}), the term on the second line describes the confined cavity electromagnetic field as a set of harmonic oscillators, where $\hat{b}_\alpha^\dagger$ and $\hat{b}_\alpha$ are the bosonic creation and annihilation operators associated with the photonic mode indexed ``$\alpha$'', and $\omega_{\alpha}$ denotes the corresponding cavity frequency.

\begin{figure*} 
    \centering
    \includegraphics[width=17cm]{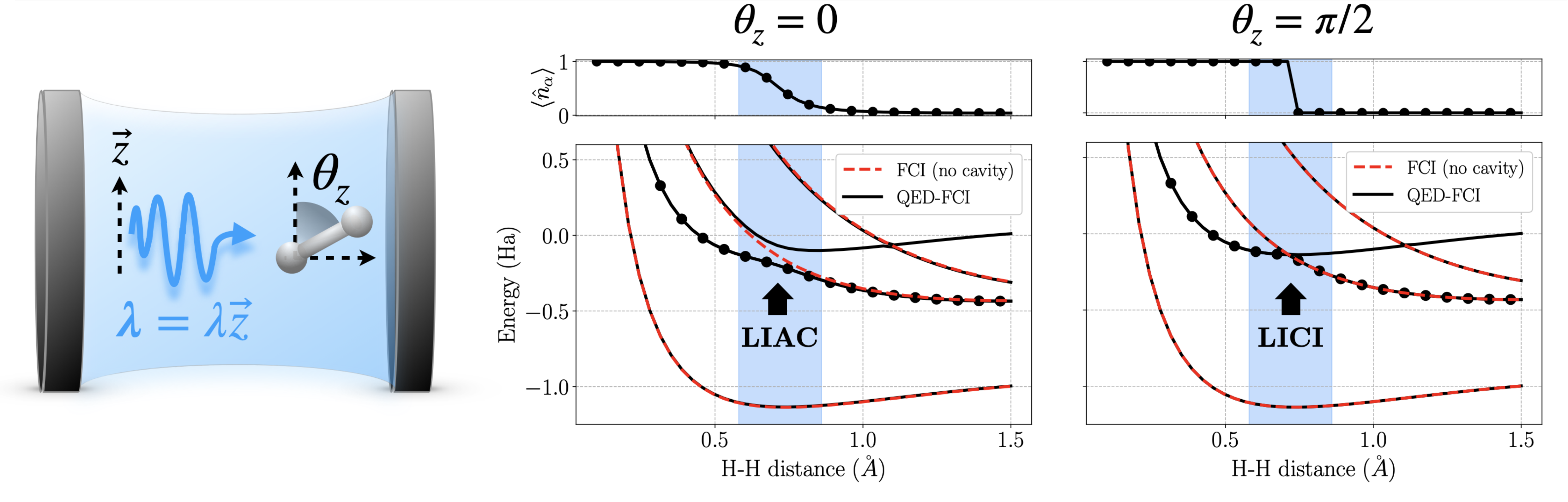}   
\caption{
\textbf{ Polaritonic Potential Energy Surfaces of H$_2$ in cavity showing Light-Induced Avoided Crossings and Conical Intersections.} \linebreak
The  molecule is described with the STO-3G basis and coupled to a single cavity mode with three photons maximum and $ \lambda =0.08$, $  \omega = 1$~Ha. 
All the  FCI and QED-FCI results presented here were obtained using the \textit{QuantNBody} package~\cite{yalouz2022quantnbody}. 
\textbf{Left panel:} Schematic representation of the system, where the quantized electromagnetic field is aligned along the $z$-axis, with coupling strength $\lambda$ and molecular tilt angle $\theta_z$ relative to the field direction.  
\textbf{Middle panel:} PES for $\theta_z = 0$ when the molecule is aligned with the field.
Black curves correspond to QED-FCI results (including cavity effects), while red dashed curves represent the FCI reference (without cavity).  
As shown, a LIAC (\textit{i.e.} a Rabi splitting) appears between the first and second polaritonic excited states for $\theta_z = 0$.
In this case, the photon number in the first polaritonic excited state (see black dotted lines in top panel) smoothly decreases from one to zero.  
\textbf{Right panel:} Same as the middle panel but for $\theta_z = \pi/2$, when the molecule is orthogonal to the field.  
For $\theta_z = \pi/2$, the LIAC transitions into a LICI, marked by a strict degeneracy between the first and second excited polaritonic states and a sudden photon-number change from one to zero (see top panel on the right). 
All these results highlight the strong dependence of the field/molecular orientations on the arising of cavity-induced effects~\cite{moiseyev2008laser,vsindelka2011strong}. }
    \label{fig:H2_illustration}
\end{figure*}

Finally, the remaining terms in the last line of Eq.~(\ref{eq:Pauli_fierz_ham}) correspond to the electron-photon interaction Hamiltonian. 
These terms describe the electron-photon coupling through photonic absorption/emission processes, as well as the self-dipole terms that affect the electronic structure. 
The vector $\boldsymbol{\lambda}_\alpha$ encodes the light-matter coupling amplitudes (in the $\vec{x}$, $\vec{y}$, and $\vec{z}$ directions) between the electron and the photonic mode $\alpha$, while the molecular dipole operator $\bold{d}$ is defined as~\cite{haugland2020coupled,el2024toward}:
\begin{equation}
\bold{\hat{d}} = \sum_{pq} \bold{d}_{pq} \hat{E}_{pq}, \quad \text{with} \quad \bold{d}_{pq} = \bold{d}_{pq}^{e} + \delta_{pq} \frac{\bold{d}^{nuc}}{N_e},    
\end{equation}
where $\bold{d}_{pq}^{e}$ are the electronic dipole integrals (in the directions $\vec{x}$, $\vec{y}$, and $\vec{z}$), and $\bold{d}^{nuc}$ the nuclear dipole contributions which remain constant at a fixed geometry (within the Born-Oppenheimer approximation).

Solving the \textit{ab initio} polaritonic chemistry problem reduces to solving the stationary Schrödinger equation for the Pauli-Fierz Hamiltonian, which governs the coupled fermion-boson system:
\begin{eqnarray} \label{eq:Schrodinger}
\hat{H} \ket{\Psi_k} = E_k \ket{\Psi_k}, 
\end{eqnarray}
where $\ket{\Psi_k}$ represents the $k$-th polaritonic eigenstate with energy $E_k$.
In this work, we consider a single cavity mode $\alpha$, allowing each polaritonic eigenstate to be exactly expressed within the so-called ``QED'' Full Configuration Interaction (QED-FCI) framework as:
\begin{equation} \label{eq:eigenstates}
\ket{\Psi_k} = \sum_I \sum_{n_\alpha} C_{I,n_\alpha} \ket{\Phi_I} \otimes \ket{n_\alpha}, 
\end{equation}
where $\ket{\Phi_I}$ denotes the electronic many-body basis states (\textit{i.e.} a Slater determinant), and $\ket{n_\alpha}$ represents the Fock state associated with the (single) cavity mode denoted $\alpha$, characterized by $n_\alpha$ photons.

\subsection{ Light-induced Avoided Crossing and Conical intersection }

Solving the Schrödinger equation in Eq.~(\ref{eq:Schrodinger}) provides direct access to the polaritonic potential energy surfaces (PES) of a molecular system confined within a cavity. Compared to molecules in \textit{vacuum}, these {``dressed''} PES can be significantly altered due to strong light-matter interactions, leading to exotic features such as Light-Induced Avoided Crossings (LIACs) and Light-Induced Conical Intersections (LICIs).



Trying to accurately describe these polaritonic effects will represent a central goal of our quantum computing development. 
Therefore, we will here briefly provide some physical insight into these light-induced phenomena, focusing on our testbed system: the H$_2$ molecule coupled to a single photonic mode inside an optical cavity.
In Figure~\ref{fig:H2_illustration} we provide an overview of the system and the impact of cavity interactions on the PES. The left panel depicts the theoretical setup.
Here, we consider the quantized electromagnetic field fully aligned along the $z$-axis, while the molecular system H$_2$  presents a tilt angle $\theta_z$ relative to the field.
The light-matter interactions are characterized by the coupling strength $\lambda = 0.08$ and the cavity frequency $\omega = 1$ Ha, designed to match the vertical excitation energy between the ground and the first excited state of the bare molecule around the Franck-Condon point.
By varying $\theta_z$, we show how the formation of hybrid light-matter states affects the resulting polaritonic PES.
All FCI and QED-FCI calculations were performed using the \textit{QuantNBody} package~\cite{yalouz2022quantnbody} we recently developed.

The middle and right panels of Figure~\ref{fig:H2_illustration} compare the PES obtained without cavity interactions (red dashed curves, FCI) and with cavity effects included (black curves, QED-FCI). The middle panel corresponds to $\theta_z = 0$ (molecule aligned with the field), where strong light-matter coupling induces a LIAC between the first and second excited polaritonic states, manifesting as a Rabi splitting (see blue-shaded area).
This feature is accompanied by a smooth transition in the photon number of the first excited polaritonic state, gradually decreasing from one to zero over the avoided crossing (see the top panel of the middle figure).
When the molecular orientation is changed to $\theta_z = \pi/2$, this LIAC evolves into a LICI, characterized by a strict degeneracy between the first and second excited polaritonic states and a sudden change in photon number from one to zero (see the top panel on the right). 
This transition highlights the strong dependence of cavity-induced effects on molecular orientation and demonstrates how light-matter coupling can fundamentally reshape molecular PES.

\begin{figure}[t]
    \centering
    \includegraphics[width=0.85\columnwidth]{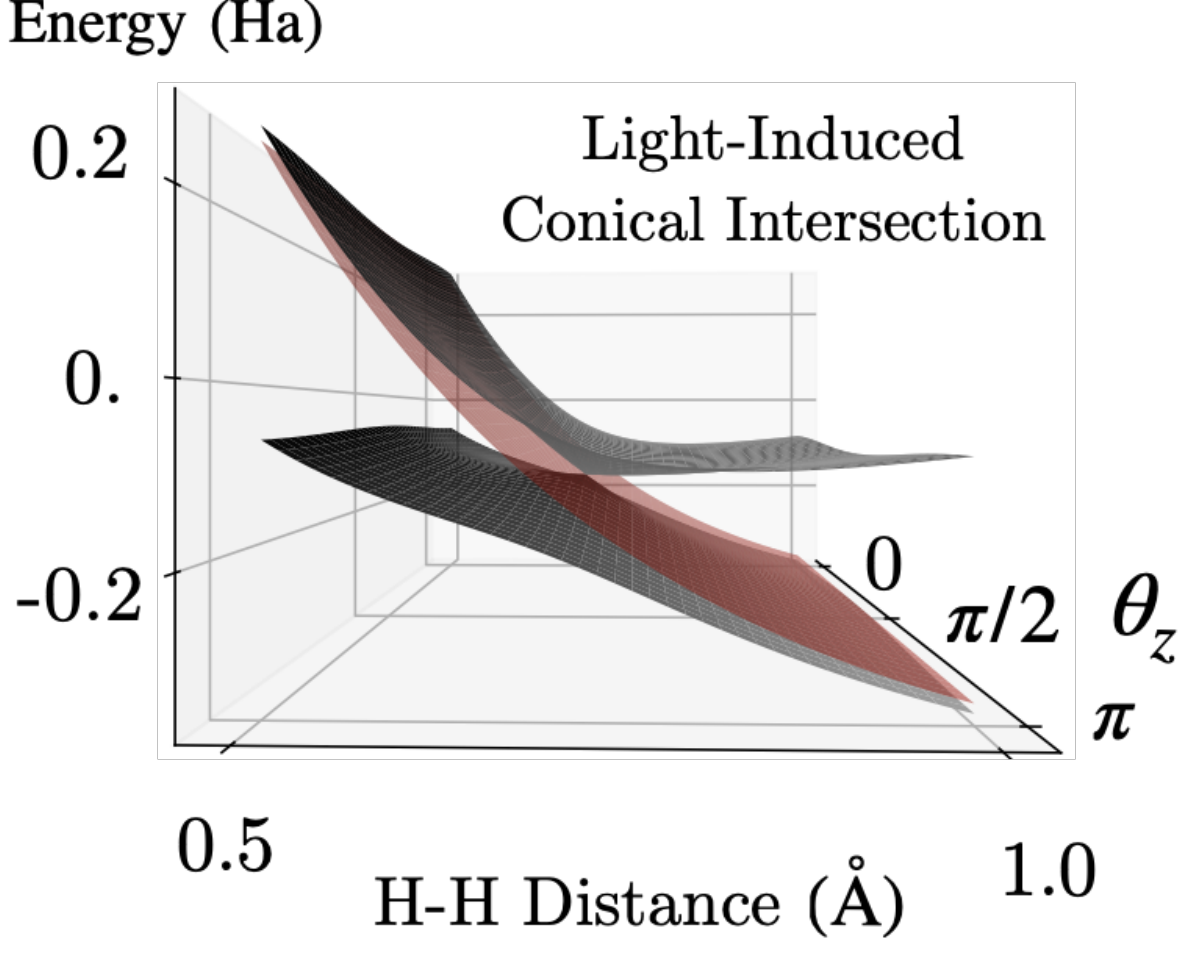}
    \caption{ 
\textbf{ Light-Induced Conical Intersection for H$_2$ in a cavity.} \linebreak 
 PES of the first and second polaritonic eigenstates for H$_2$ (minimal STO-3G basis set) in a cavity with a single mode containing at most three photons, with coupling strength $\lambda = 0.08$, cavity frequency $\omega = 1$ Ha.  
Black surfaces represent the polaritonic  PES, while the red surface corresponds to the first excited state of the bare molecule.  
Results obtained using the \textit{QuantNBody} package~\cite{yalouz2022quantnbody}.
}
    \label{fig:3D_LICI}
\end{figure}

From a theoretical perspective, such a formal framework must be understood as a kind of interaction picture whereby absorption and stimulated emission are treated as direct resonances between degenerate polaritonic states (roughly: ground-state molecule plus one photon and excited-state molecule with no photon). 
LIACs and LICIs can compete with intrinsic conical intersections present in the absence of a cavity and reshape the PES, thereby enabling controlled modifications of chemical reactivity~\cite{csehi2017lici}.
They also are responsible of nonadiabatic-like phenomena that  can profoundly influence spectroscopic properties, such as vibronic intensity borrowing in polyatomic molecules~\cite{fabri2020lici-a,fabri2020lici-b}. 
{The fundamental reason for this is that the cavity field polarization creates an actual quantization axis that changes the status of the overall rotational degrees of freedom (external and isotropically free for an isolated molecule). 
In particular, for a diatomic system  like H$_2$, the angle between the internuclear axis and the polarized field (here $\theta_z$) turns into an internal degree of freedom with respect to the polaritonic system. 
This changes the internal dimensionality of the system from one to two, thus creating the opportunity for a LICI of codimension two to occur between two dressed PES.
We provide in Fig.~\ref{fig:3D_LICI} a three-dimensional representation of the LICI occurring between the first and second polaritonic excited states of H$_2$ against both the interatomic distance and the angle $\theta_z$.
The case of a LIAC is then a less restrictive situation, close to a LICI point, but whereby the value of the relative angle occurs to be such that the coupling there is nonzero, much as in field-free situations for which a ``natural'' conical intersection is surrounded by avoided crossings.}

 \section{ Quantum Computing Background }
\label{sec:III}
 
Developing computational strategies for accurately describing and predicting the emergence of LIACs and LICIs in the spectrum of polaritonic systems is crucial to advance our understanding of novel cavity-induced chemical phenomena.
Given the inherently quantum nature of polaritonic systems, quantum computing provides a natural and promising framework for tackling such problems, provided that two central questions can be addressed.
First, how can we efficiently map and solve this problem on quantum computers in a way that enables the accurate description of excited states?
Second, which quantum computing architectures, including \textit{qubit-based}, \textit{qudit-based}, or \textit{hybrid qubit-qumode} platforms, would be the best suited to perform such calculations efficiently?

In this section, we first introduce the reference quantum algorithm used to provide a balanced description of multiple polaritonic states, namely: the State-Averaged Variational Quantum Eigensolver (SA-VQE).
We then examine how the electron-photon problem can be mapped onto the three different types of quantum platforms considered.
Finally, we present a distinct quantum circuit ans\"{a}tze tailored to each platform, specifically designed to encode strongly correlated light-matter states within the SA-VQE framework.

 \subsection{ State-Averaged Variational Quantum Eigensolver }
\label{sec:SAVQE}
 The SA-VQE method, also referred to as Subspace-Search~\cite{nakanishi2019subspace} or Multi-State Contracted VQE~\cite{parrish2019quantum}, defines a type of algorithm recently introduced to extend beyond ground-state calculations and address low-lying excited states within the constraints of Noisy Intermediate-Scale Quantum (NISQ) devices.  
Unlike standard VQE~\cite{peruzzo2014variational}, which targets only the ground state, SA-VQE is designed to provide a collective description of a set of $N_S$ low-lying states, which makes it particularly well suited for capturing close degeneracies and avoided-crossings. 
Such approaches, along with their extensions, have proven especially effective in providing accurate descriptions of conical intersections in photochemical systems~\cite{yalouz2021state,yalouz2022analytical,illesova2025transformation,beseda2024state,omiya2022analytical}.

From a theoretical perspective, the goal of the SA-VQE algorithm is to generate an ensemble of trial states that will be optimized to encode the first $N_S$ low-lying eigenstates of a quantum system using a hybrid quantum-classical approach. 
The procedure starts by selecting a set of $N_S$ orthogonal input states $|\psi_k\rangle$, where $k = 1, \ldots, N_S$, which are easy to prepare on a reference quantum platform. 
Each of these input states is then transformed into parametrized trial states $|\psi_k(\vec{\theta})\rangle$ according to  
\begin{equation}
|\psi_k(\vec{\theta})\rangle = \hat{U}(\vec{\theta}) |\psi_k\rangle,
\end{equation}
where $\hat{U}(\vec{\theta})$ is a parameterized unitary operator encoded in a single "ansatz" quantum circuit with quantum gate parameters $\vec{\theta}$, which will be optimized during the course of the algorithm. The resulting trial states serve as approximations of the exact eigenstates $|\Psi_k\rangle$ of a reference system with Hamiltonian $\hat{H}$. 
To achieve convergence towards the targeted eigen-space, the algorithm minimizes the state-averaged energy over the set of trial states:
\begin{equation}\label{eq:SA-VQE-energy}
E^\text{SA-VQE} (\vec{\theta}) = \frac{1}{N_S}  \sum_{k=1}^{N_S}  \langle \psi_k(\vec{\theta}) | \hat{H} | \psi_k(\vec{\theta}) \rangle.
\end{equation}
In practice, energy measurements are performed at the end of the ansatz quantum circuit for each trial state.
The resulting state-averaged energy is then computed on a classical computer, which also performs optimization of the variational parameters $\vec{\theta}$. 
According to the ensemble Rayleigh-Ritz variational principle~\cite{gross1988rayleigh}, minimizing the state-averaged energy ensures that the optimized trial states $|\psi_k(\vec{\theta}^*)\rangle$ (with $\vec\theta^*$ the optimized circuit parameters) converge toward the exact low-lying eigen-subspace of the target Hamiltonian $\hat{H}$.
To complete the calculation, a subspace resolution step is required to extract the individual adiabatic eigenstates and energies.
Several methods have been proposed for this purpose, for example, using an additional state-specific circuit optimization (see, for example, Refs.\cite{nakanishi2019subspace,yalouz2022analytical}).
In the present work, we choose to adopt another approach as introduced by Parrish et al.~\cite{parrish2019quantum}, which involves the classical construction and diagonalization of the Hamiltonian matrix sub-block of size $N_S \times N_S$ defined by
\begin{equation}
    \bold{H} = \sum_{k,k'=1}^{N_S} \bold{H}_{kk'}  |\psi_k(\vec\theta^*) \rangle \langle \psi_{k'}(\vec\theta^*)|,
\end{equation}
with matrix elements $\bold{H}_{kk'}  = \langle \psi_k(\vec{\theta}) | \hat{H} | \psi_{k'}(\vec{\theta}) \rangle$ formed by the converged trial states. 
For practical details on how to build and estimate these matrix elements, we refer the interested reader to Refs.~\cite{parrish2019quantum} and ~\cite{nakanishi2019subspace} where strategies have been introduced and thoroughly discussed. 
To conclude this introductory section on SA-VQE, we will briefly mention the two practical NISQ-friendly features brought about by this algorithm.
First, it significantly reduces the quantum resources required by simultaneously optimizing multiple states within the same quantum circuit. This is particularly beneficial for NISQ devices, where resource limitations are a major concern. Second, it guarantees that the generated trial states remain orthogonal throughout the optimization process. Specifically, if the input states $|\psi_k\rangle$ are orthogonal, which means $\langle \psi_k | \psi_{k'} \rangle = \delta_{kk'}$, then the following condition is maintained throughout the optimization:
\begin{equation}
\begin{split}
    \langle \psi_k(\vec{\theta}) | \psi_{k'}(\vec{\theta}) \rangle &= \langle \psi_k | \hat{U}(\vec{\theta})^\dagger \hat{U}(\vec{\theta}) | \psi_{k'} \rangle \\
    &= \langle \psi_k | \psi_{k'} \rangle  = \delta_{kk'},
\end{split}
\end{equation}
as we are here working with a unitary transformation $\hat{U}(\vec{\theta})^\dagger \hat{U}(\vec{\theta}) = \hat{U}(\vec{\theta}) \hat{U}(\vec{\theta})^\dagger = \bold{1}$.


\subsection{ Platform-Specific Mappings for Fermions and Bosons: \\ 
Methods for Qubits, Qudits, and Qumodes
}

Starting from the SA-VQE method, we now discuss the \textit{ab initio} polaritonic chemistry problem across three distinct quantum platforms: \textit{(i)}~\textit{qubit-based}, \textit{(ii)}~\textit{qudit-based}, and \textit{(iii)}~hybrid \textit{qubit-qumode} hardware.  
To proceed, we first need to address the key question of how to simultaneously map electronic and photonic degrees of freedom onto these different types of quantum information units.
While various strategies could be considered, in this work we will adopt an approach that ensures a fair and consistent basis for future cross-platform comparisons.

\subsubsection{ Fermionic Mapping: \\ General Strategy Across Platforms  }
\label{sec:JW_mapping}

To ensure a consistent and fair starting point in our future circuit development, we made a specific choice: adopting the same type of mapping for the electronic degrees of freedom across all three platforms.
Indeed, it should be noted that all platforms considered here offer the possibility to  manipulate two-level systems. 
This includes \textit{qudit-based} architectures, where certain qudits can be reduced to ``effective'' qubits (by setting the dimension \(d = 2\)), enabling operations governed by the conventional Pauli algebra. 
This feature is particularly relevant, as fermionic degrees of freedom, when expressed in second quantization, naturally map onto quantum information units that behave like two-level systems (\textit{i.e.}, qubit-like).
As a consequence, for all three hardware platforms, we choose to adopt the Jordan-Wigner transformation to map each fermionic spin orbital of the polaritonic problem onto qubit-like units of information. 
In this mapping, the occupancy of a spin orbital is directly represented by the state of a single qubit(-like) system: a spin orbital occupied by one electron corresponds to the qubit state $|1\rangle$, while an unoccupied orbital corresponds to $|0\rangle$.
Within this framework, fermionic second-quantized operators are mapped onto Pauli operators according to the following rules:
\begin{equation}
\label{eq:Jordan_Wigner}
    \hat{a}_p^\dagger \xrightarrow{} \left( \bigotimes_{q=1}^{p-1} \hat Z_q \right) \otimes \frac{\hat X_p - i \hat Y_p}{2},
\end{equation}
where \( \hat X_j\), \( \hat Y_j\), and \( \hat Z_j\) are the Pauli matrices associated with the \(j\)-th two-level system defined such as
\begin{equation}
\begin{split}
    \hat {X}   &= \ketbra{0}{1} + \ketbra{1}{0}, \\
     \hat {Y} &= -i \left( \ketbra{0}{1} - \ketbra{1}{0} \right),\\
     \hat {Z} &=   \ketbra{0}{0} - \ketbra{1}{1} .\\
\end{split}
\end{equation}
Note that these operators naturally integrate into the more generalized qudit algebra (for the specific case of \(d=2\)), which will be detailed later.

\subsubsection{ Bosonic Mapping: \\ Platform-Specific Strategy }

Turning now to bosonic degrees of freedom (i.e. cavity mode), we will now discuss our platform-adapted strategy for mapping them onto the different quantum platforms considered. 
A key distinction between bosons and fermions lies in the fact that bosons can occupy the same quantum mode in unlimited numbers, whereas fermions are constrained by the Pauli exclusion principle. 
This fundamental difference makes the mapping of bosonic degrees of freedom onto qubit-based quantum hardware inherently more challenging, often requiring a truncation of the bosonic Fock space to a finite-dimensional representation (although, as we will see later, this is not always necessary if we consider  \textit{qumodes}). 
In what follows, we detail the specific approaches envisioned for each of the three platforms.

  \vspace{0.25cm}
  
\paragraph{{Qubit-Based Architectures.}}
Mapping bosonic modes onto qubits is theoretically feasible but requires significantly more quantum resources than mapping fermionic modes (see Refs.~\cite{somma2002simulating, veis2016quantum, peng2023quantum, knapik2025quantum}). In this work, we consider a ``direct-boson mapping'', the action of which can be summarized as follows.
For a single photonic mode, we represent the occupancy using $N_B^\text{max}+1$ qubits, where $N_B^\text{max}$ denotes the maximum number of bosons that can occupy the mode. 
Specifically, we focus on bitstrings in which exactly one bit is set to ``1'', and its position indicates the number of bosons present.
For instance, if $N_B^\text{max} = 2$, the bosonic vacuum is encoded as the three-qubit state \(\ket{0}_B \rightarrow \ket{100}\), the state with one boson as \(\ket{1}_B \rightarrow \ket{010}\), and the state with two bosons as \(\ket{2}_B \rightarrow \ket{001}\). The extension of this mapping to multiple modes and higher boson numbers is straightforward.
An interesting feature of this mapping is that it enables a direct expression of bosonic creation and annihilation operators in terms of qubit operators. 
Specifically, the bosonic creation operator is mapped as follows:
\begin{align}\label{eq:DBM_qubits}
    \hat{b}^{\dagger} &\xrightarrow{}  \sum_{q=0}^{N_B^\text{max}-1} \sqrt{q+1} \, \hat \sigma^{+}_{q} \hat \sigma^{-}_{q+1},
\end{align}
where $ \hat \sigma^{\pm}_q = (\hat X_q \mp i \hat Y_q)/2 $ are the qubit raising and lowering operators for the \(q\)-th qubit.

 \vspace{0.25cm}

\paragraph{{Qudit-Based Architectures.}} From a computational perspective, qudits are \textit{discrete quantum variables} that obey the so-called Gell-Mann algebra, which generalizes the Pauli algebra to systems with dimensionality \(d > 2\)~\cite{bertlmann2008bloch, kimura2003bloch}. \linebreak
Although mappings between bosonic degrees of freedom and finite-dimensional systems have been discussed in some works (see for example Ref.~\cite{batista2004algebraic}), an explicit formulation of a boson-to-qudit operator mapping does not appear to be readily available in the existing quantum computing literature.
Therefore, we present here an approach for mapping bosonic degrees of freedom onto qudits, which naturally takes advantage of the local \(d\)-level systems by employing the Generalized Gell-Mann Matrices (GGM). 
In the case of a qudit with $d$ levels, the set of GGMs typically includes a total of  $d(d-1)$  symmetric and antisymmetric matrices, denoted respectively $\hat \Lambda_{ll'}^{X}$ and $\hat \Lambda_{ll'}^{Y}$, which are defined as follows (see Refs.~\cite{bertlmann2008bloch, kimura2003bloch}):
\begin{equation}
\begin{split}
    \hat \Lambda_{ll'}^{X}   &= \ketbra{l}{l'} + \ketbra{l'}{l}, \textcolor{black}{ \quad \text{for} \quad {0 \leq l < l' \leq d-1}} \\
    \hat \Lambda_{ll'}^{Y} &= -i \left( \ketbra{l}{l'} - \ketbra{l'}{l} \right), \textcolor{black}{ \quad \text{for} \quad {0 \leq l < l' \leq d-1}}
\end{split}
\end{equation}
where \(\ket{l}\) (with \(l = 0, \ldots, d-1\)) represents the intra-level states of the qudit.
Note that the superscript notation is used here to indicate that these matrices act analogously to the conventional Pauli 
$X$ and $Y$ gates, but only upon specific sublevels of the reference qudit.
For the particular case $d = 2$, the GGM turn back to the Pauli matrices.
Let us now focus on the two subsets of matrices that connect consecutive levels, \textit{i.e.} $l' = l + 1$. 
From these matrices we can derive the following property:
\begin{equation}
   \frac{1}{2}( \hat \Lambda_{ll+1}^{X} - i  \hat \Lambda_{ll+1}^{Y}) = \ketbra{l+1}{l},
\end{equation}
which allows us to introduce a direct mapping for bosonic degrees of freedom onto a single $d$-level qudit, as follows:
\begin{equation}\label{eq:DBM_qudit}
   \hat b^\dagger \rightarrow \sum_{l=0}^{N_B^\text{max}-1} \frac{\sqrt{l+1}}{2} \left(\hat  \Lambda_{ll+1}^{X} - i  \hat \Lambda_{ll+1}^{Y} \right),
\end{equation}
This mapping establishes a direct correspondence between the qudit level and the number of bosonic particles within a single mode. 
In this framework, the zeroth qudit level corresponds to the bosonic vacuum, the first qudit level denotes the state comprising one boson, and so on, up to the $(d-1) = N_B^\text{max} $ qudit level, which represents the presence of $N_B^\text{max}$ bosons. 
Consequently, the dimensionality of the qudit defines a truncated Fock space for a single mode that describes up to $N_B^\text{max}$ bosons.

 \vspace{0.25cm}

\paragraph{{Hybrid Qubit-Qumode-Based Architectures.}}
Within these emerging hardware platforms~\cite{eickbusch2022fast,crane2024hybrid}, qumodes offer a natural and efficient way to encode bosonic degrees of freedom, as they directly correspond to quantum harmonic oscillators with associated creation and annihilation operators, $b^\dagger_\text{qumode}$ and $b_\text{qumode}$~\cite{kalajdzievski2018continuous,yalouz2021encoding}. 
These \textit{continuous variable} units of information theoretically reside in a Fock space of infinite dimension ($\dim = +\infty$), allowing the representation and exploration of higher bosonic excitations.
This property closely mirrors the physical nature of photonic modes in polaritonic systems, which are also modeled by infinite-dimensional Hilbert spaces.
This distinction becomes particularly significant when compared to discrete quantum systems such as \textit{qubits} or \textit{qudits}, which are confined to finite-dimensional state spaces (with respective dimensions $\dim = 2$ and $\dim = d$), and therefore inherently misrepresent the physical continuous nature of realistic photonic degrees of freedom.

Following recent developments~\cite{yalouz2021encoding}, we adopt a direct analog mapping between the photonic degrees of freedom in the polaritonic problem and the qumodes naturally present in hybrid quantum platforms. 
Within this framework, the photonic Fock space of a cavity mode is directly mapped onto the bosonic Fock space of a single qumode by assuming the correspondence
\begin{equation}
 \hat    b^\dagger \rightarrow  \hat   b^\dagger_\text{qumode}.
\end{equation}
Beyond the physical motivations underpinning this mapping, as we show later, this approach located at the interface of digital and analog quantum computing also offers practical advantages for implementing electron-photon entanglement using hybrid qubit-qumode gates within quantum circuits~\cite{crane2024hybrid,liu2024hybrid}.

\subsection{ Design of Compact Quantum Circuit Ans\"{a}tze   }
\label{sec:CompactAnsatz}

Building upon the previously defined fermion and boson mappings, we now focus on the design of quantum circuits to solve the polaritonic problem using the SA-VQE algorithm.
In practice, a variety of strategies can be employed, ranging from hardware efficient circuits, compact designs particularly suited for NISQ devices, to more physically motivated ansatze, such as the Unitary Coupled Cluster (UCC) ansatz~\cite{anand2022quantum}.
While much of the research on quantum circuit ansatze has focused on pure electronic structure problems, relatively little has been done in the context of \textit{ab initio} polaritonic chemistry.
A significant contribution in this area is the development of the polaritonic UCC ansatz, which has demonstrated excellent accuracy for ground states~\cite{pavosevic2021polaritonic, hassan2024simulating}. 
However, a well-known limitation of UCC based approaches is the substantial circuit depth required for their implementation, which poses challenges for near-term quantum devices. 
The computational cost increases further when targeting multiple excited states (such as in our case), necessitating the use of a generalized version of the UCC ansatz~\cite{greene2021generalized, lee2018generalized, anand2022quantum, yalouz2021state}, or extended equation-of-motion theories~\cite{pavosevic2021polaritonic}.

\begin{figure*}
    \centering
    \includegraphics[width=18cm]{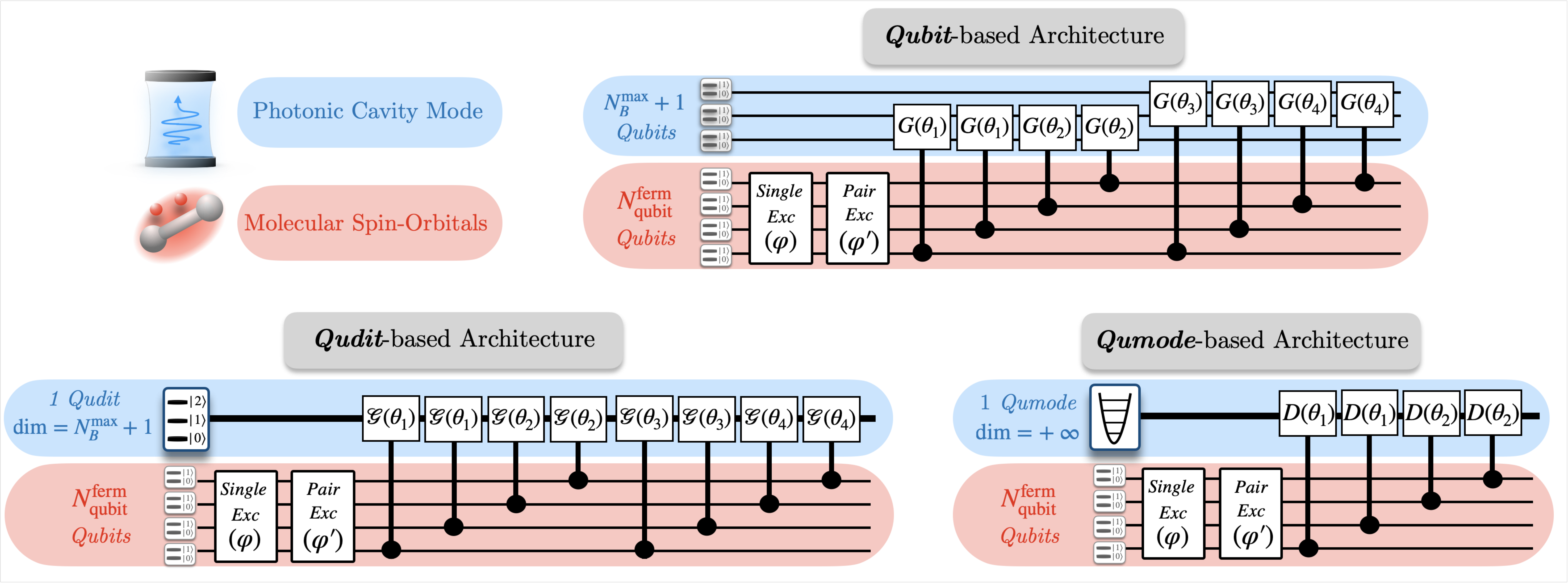}
    \caption{
    \textbf{Different quantum circuit designs proposed for encoding polaritonic eigenstates on qubit, qudit and qumode platforms. \linebreak}
    For each platform, a \textit{single layer} of the quantum circuit ansatz is shown with red areas highlighting the parts dedicated to the molecular electronic structure, while blue areas correspond to the photonic cavity mode.  
Regardless of the platform, qubit-like degrees of freedom are used to encode the electronic structure based on a single layer of the Gate-Fabric ansatz~\cite{anselmetti2021local} composed of a single-excitation fermionic gate followed by a pair-excitation gate.  
The electron-photon entanglement, however, is encoded in a platform-specific manner.
In the circuit setup, for a given pair of qubits associated with the same spatial orbital we employ the same type of controlled gate with a shared parameter to preserve spin symmetries  $\hat{S}^2$  and $\hat{S}_z$ (see proof in Appendix~\ref{app:proof_symmetry}).
Note in this illustrative case that for both qubit and qudit platforms which can only encode a discrete bosonic dimension we consider  a maximum number of $N_B^\text{max} = 2$ photons.
\textbf{Top panel:} Qubit-based architecture carries interleaved layers of controlled-Givens gate $CG(\theta)$ (see Eq.(\ref{eq:CGiven})) which will generate photonic excitation/de-excitation depending on the occupancy of a same spatial orbital. 
\textbf{Lower left panel:} Qudit-based architecture where similar approach is carried but here with controlled-gate $C \cal G(\theta)$ (see Eq.(\ref{eq:Control_givens_qudit})) propagating photonic excitations between the internal sub-levels of the qudit.
\textbf{Lower right panel:} Qumode-based architecture where we use  controlled displacement gates $CD(\theta)$ (see Eq.(\ref{eq:CD_gate})) which can (in principle) allow propagation through the infinite qumode Fock space.
    }
    \label{fig:Compact_circuits}
\end{figure*}

To address these challenges, we will now introduce the design of compact and versatile quantum circuit ans\"{a}tze that strike a balance between physical motivation and hardware efficiency, tailored to each of the platforms considered. \linebreak
Since electronic degrees of freedom are mapped similarly onto local two-level systems across all platforms, we choose to begin with a well-established and reliable approach for treating the purely electronic structure: the so-called Gate-Fabric Ansatz (see Ref.~\cite{anselmetti2021local}). 
Building on this common foundation, our circuit design then diverges by leveraging the unique features of the three distinct quantum computing architectures considered particularly in the way electron-photon entanglement is encoded. 
In this manner, the new circuits developed here can be viewed as extended Fermion-Boson Gate-Fabric Ans\"{a}tze, adapted to different types of quantum computing platforms. Figure~\ref{fig:Compact_circuits} illustrates the various quantum circuit ans\"{a}tz architectures considered in this work, with red and blue shaded areas indicating the regions dedicated to fermionic and bosonic degrees of freedom, respectively.
    
\subsubsection{Fermionic Gate Fabric Ansatz:\\ A Generalized Strategy Across Platforms }
 
The ``Gate-Fabric Ansatz'' introduced in Ref.~\cite{anselmetti2021local} belongs to a class of circuits that stand at the crossroads between hardware-efficient approaches and physically motivated ones. 
It is based on fundamental building blocks that encode local fermionic interactions while preserving key symmetries, such as spin (\(\hat{S}^2\) and \(\hat{S}_z\)) and the total particle number. 
Under the Jordan-Wigner mapping (as defined in Sec.~\ref{sec:JW_mapping}), two four-qubit unitary transformations can be defined.  
The first transformation, denoted \(\hat{U}^{\rm pair}(\theta)\), encodes local double-electron excitation
while the second transformation, denoted \(\hat{U}^{\rm single}(\theta)\), implements local spin-free single-electron excitation (\textit{i.e.}, a spatial orbital rotation).
In practice, building an ansatz circuit based on the Gate-Fabric philosophy relies on the repeated application of these parametrized excitation operators to the qubits encoding the electronic structure.  
The circuit length can be adjusted at will in the same way as in hardware-efficient ansätze to enhance encoding flexibility.  
However, in this case, the local building block retains a clear physical meaning and it has been demonstrated to be highly expressive in its description of strongly correlated states in many molecular systems (without cavity)~\cite{anselmetti2021local}.  
Interestingly, as explained in Refs.~\cite{anselmetti2021local,arrazola2022universal}, the mathematical foundations underpinning both double- and single-electronic excitation operations 
are based on the concept of ``\textit{Givens rotations}'' \(\hat{G}(\theta)\) between two qubits $q$ and $q'$, as defined by  
\begin{equation}
\begin{split}
    \hat{G}_{qq'}(\theta) &= \exp\left(-i\frac{\theta}{2} (\hat X_q \otimes \hat Y_{q'}  - \hat Y_{q} \otimes \hat X_{q'} )\right)  
\end{split}
\end{equation} 
which act as follows
\begin{equation}
\label{eq:givens_rotation}
\begin{split}
    \hat{G}_{qq'}(\theta) \ket{0_{q}1_{q'}} & = \cos(\theta) \ket{0_{q}1_{q'}} + \sin(\theta) \ket{1_{q}0_{q'}} , \\
    \hat{G}_{qq'}(\theta) \ket{1_{q}0_{q'}} & = \cos(\theta) \ket{1_{q}0_{q'}} - \sin(\theta) \ket{0_{q}1_{q'}} ,
\end{split}
\end{equation} 
while leaving any other state unchanged.
We will show in the following sections that such elementary operations can also be used to design fermion-boson entangling gates.

\begin{figure}[t]
    \centering
    \includegraphics[width=\columnwidth]{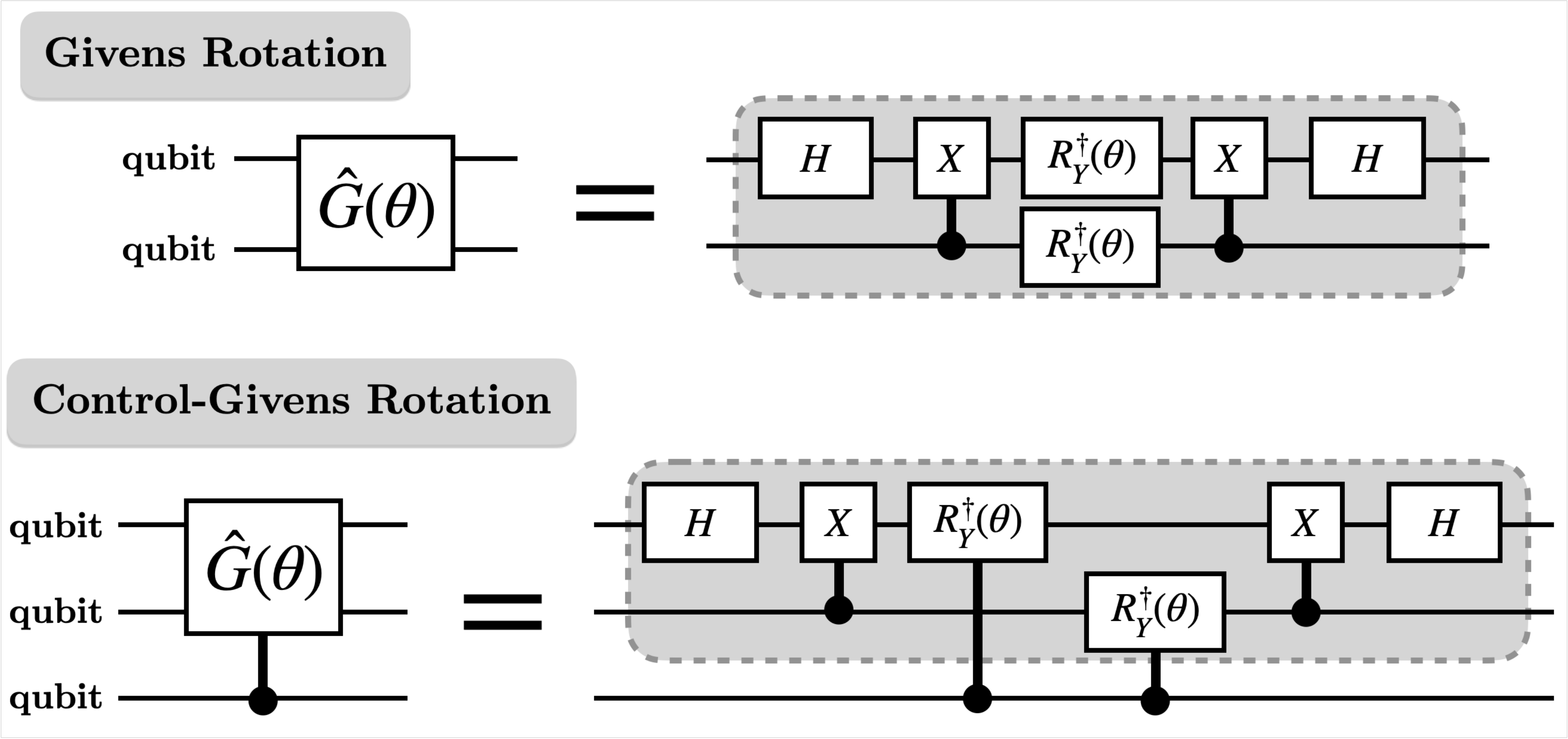}
    \caption{\textbf{Quantum circuits for implementing a Givens rotation (upper panel) and a controlled-Givens rotation (lower panel)}.
    The controlled-Givens rotation requires only four entangling control gates due to circuit symmetry.
    This symmetry enables the cancellation of certain gates when the control qubit is in the zero state, resulting in a more compact representation.
    Note that alternative implementations based on the decomposition of exponentiated Pauli-strings are possible, but they generally result in slightly deeper circuits (see Appendix~\ref{appendix:alternative circuit} where we discuss such alternatives).
    }
    \label{fig:Givens_circuits}
\end{figure}

\subsubsection{ Fermion-Boson Entanglement Gates:\\   Platform-Specific Strategy }
\label{sec:Entangling gates}

Going beyond the bare electronic structure problem, we now aim to extend the conventional fermionic Gate-Fabric Ansatz to the polaritonic framework. 
This requires incorporating operations capable of capturing light-matter entanglement, which is essential for accurately modeling states near a LIAC region.
To develop quantum circuit ansätze that are both compact and physically motivated, in the following we present the hardware-adapted strategies explored for each of the three platforms.

  \vspace{0.25cm}
  
\paragraph{{Qubit-Based Architectures.}}
Assuming the use of the direct boson mapping introduced in Eq.~(\ref{eq:DBM_qubits}), electron-photon entanglement is incorporated through a set of controlled-{Givens} rotations:  
\begin{equation}\label{eq:CGiven}
    C\hat{G}(\theta) = \ketbra{0_q}{0_q} \otimes \hat{ \mathbf{1} } + \ketbra{1_q}{1_q} \otimes \hat{G}(\theta),
\end{equation}  
where \(\hat{G}(\theta)\), defined in Eq.~(\ref{eq:givens_rotation}), acts on the qubit register that encodes the bosonic degrees of freedom, while the control qubit \(q\) belongs to the fermionic register.  
Figure~\ref{fig:Givens_circuits} illustrates the quantum circuit architectures required to implement both the standard Givens rotation and its controlled version, as defined in Eq.~(\ref{eq:CGiven}).

The choice of such a controlled gate is physically motivated by the structure of the electron-photon Hamiltonian given in Eq.~(\ref{eq:Pauli_fierz_ham}) and can be understood as follows:
when a given spin-orbital carries one electron (\textit{i.e.} the associated control qubit \(q\) is in its \(\ket{1_q}\) state), an entangling light-matter interaction is enabled in the circuit, leading to parametrized photon emission or absorption as encoded in the Givens rotation.  \linebreak
Indeed, within the direct boson mapping on qubits (see Eq.~(\ref{eq:DBM_qubits})), the effect of the Givens rotation is to propagate the excitation (\textit{i.e.} the unique state \(\ket{1}\)) within the qubit register encoding the bosonic degrees of freedom (as mathematically shown in Eq.~(\ref{eq:givens_rotation})).  
As shown in the top panel of Fig.~\ref{fig:Compact_circuits}, we choose to interleave a series of such controlled Givens rotations to naturally allow access to higher bosonic excitations (i.e. propagating the unique ``one state'' to upper qubits in the bosonic register).  
In the circuit setup, for a given pair of qubits associated with the same spatial orbital (one for the up-spin and one for the down-spin), we employ the same type of controlled Givens rotation with a shared parameter.  
This choice is primarily motivated by the need to preserve spin symmetries \(\hat{S}^2\) and $\hat{S}_z$, which can be rigorously proven (we refer the interested reader to Appendix~\ref{app:proof_symmetry} where we provide such mathematical proof).

 \vspace{0.25cm}

\paragraph{{Qudit-Based Architectures.}}

\begin{figure}[t]
    \centering
    \includegraphics[width=0.5\columnwidth]{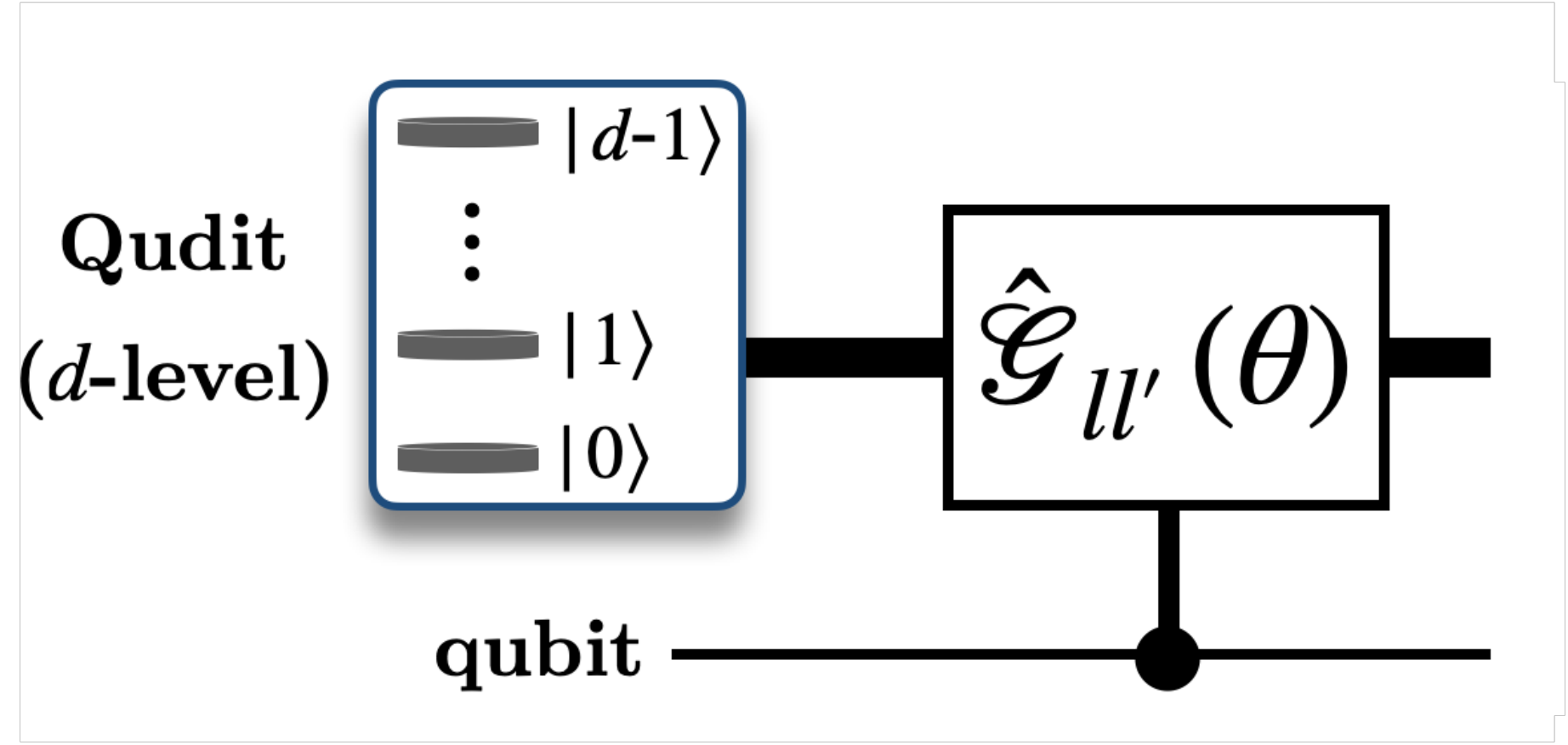}
    \caption{\textbf{Quantum circuit for a controlled-Givens rotation acting on a qudit with $d$ internal levels.} 
    This operation, which is controlled by a master qubit, implements the parametrized exponential $\hat{\mathcal{G}}_{ll'}(\theta) = \exp\left( - i\frac{\theta}{2} \hat{\Lambda}^{Y}_{ll'}  \right)  $ which performs a rotation between the two qudit sublevels indexed $l$ and $l'$. 
    }
    \label{fig:Givens_circuits_qudit}
\end{figure}

Starting from Eq.~(\ref{eq:DBM_qudit}), where we introduced a qudit-adapted version of the direct boson mapping, we follow a similar approach as in the qubit-based case to account for electron-photon entanglement \textit{via} a set of controlled operations:  
\begin{equation}\label{eq:Control_givens_qudit}
    C\hat{\mathcal{G}}(\theta) = \ketbra{0_q}{0_q} \otimes \hat{ \mathbf{1} } + \ketbra{1_q}{1_q} \otimes \hat{\mathcal{G}}(\theta),
\end{equation}
where the control qubit \(q\) resides in the fermionic register, while \(\hat{\mathcal{G}}(\theta)\) represents a Givens rotation acting on the internal levels of the qudit encoding a bosonic mode as illustrated in Figure \ref{fig:Givens_circuits_qudit}.
More explicitly, the Givens rotation in Eq.~(\ref{eq:Control_givens_qudit}) takes the form of a local $\hat{\Lambda}^{Y}_{ll+1}$-rotation between adjacent sublevels like 
\begin{equation}
\begin{split}
   \hat{\mathcal{G}}_{ll+1}(\theta) &= \exp\left( - i\frac{\theta}{2} \hat{\Lambda}^{Y}_{ll+1}  \right)  
\end{split}
\end{equation} 
which act as follows
\begin{equation}\label{eq:qudit_givens}
\begin{split}
    \hat{\mathcal{G}}_{ll+1}(\theta) \ket{l} &=  \cos(\theta) \ket{l} + \sin(\theta) \ket{l+1} , \\
    \hat{\mathcal{G}}_{ll+1}(\theta) \ket{l+1} &=  \cos(\theta) \ket{l+1} - \sin(\theta) \ket{l} ,
\end{split}
\end{equation} 
while leaving any other state unchanged. 
Here again, the choice of the controlled version of this Givens rotation defined in Eq.~(\ref{eq:Control_givens_qudit}) is physically motivated and follows the same rationale as in the previously discussed qubit-based case. 
However, in the present qudit scenario, the role of the Givens rotation \(\hat{\mathcal{G}}(\theta)\) is to mediate bosonic excitations and de-excitations within the internal sublevels of the single reference qudit. 
As illustrated in Fig.~\ref{fig:Compact_circuits}, for each pair of qubits associated with the same spatial orbital (one corresponding to up-spin and the other to down-spin), we employ a shared-parameter controlled Givens rotation to preserve spin symmetries, specifically \(\hat{S}^2\) and \(\hat{S}_z\) (see Appendix~\ref{app:proof_symmetry} for mathematical proof).

To conclude on the choice of this gate, we emphasize that it was also motivated with hardware feasibility in mind.
Indeed, controlled-qudit Givens rotations are not merely theoretical constructs, and recent studies based on optimal control theory have explored pathways for their practical implementation on real quantum hardware.
In particular, for transmon-based architectures, strategies have been proposed to realize such operations using (echoed) cross-resonance pulses with high fidelity. 
For further details, see Refs.~\cite{fischer2023universal, omanakuttan2023qudit, brennen2005criteria, bullock2005asymptotically, muthukrishnan2000multivalued} and references therein.

\begin{figure}[t!]
    \centering
    \includegraphics[width= \columnwidth]{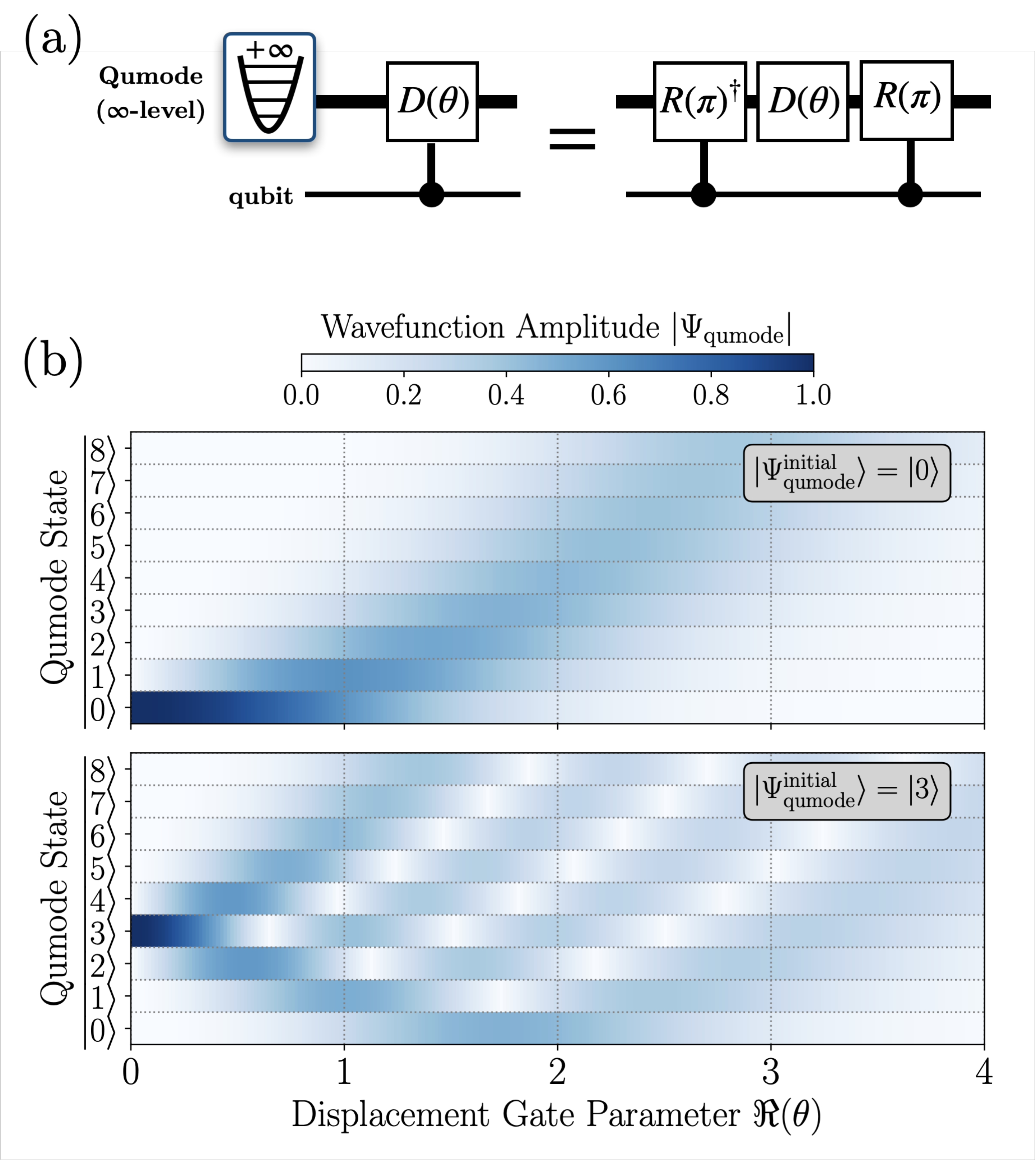}
    \caption{\textbf{Controlled-displacement gate acting on a Qumode.}
    \textbf{Top panel (a):} Quantum circuit decomposition of the operation  $CD(\theta) = \exp\left( \hat{Z}_q \otimes \left( \theta \hat{b} - \theta^* \hat{b}^\dagger \right) \right)$,  controlled by a master qubit.
    The resulting gate can be in principle decomposed with two native entangling controlled-Rotation gates $CR(\pi)$  (as explained in Ref.~\cite{crane2024hybrid}).
    \textbf{Bottom panel (b):} Illustration of the effect of the Displacement gate  on a qumode (as defined in Eq.(\ref{eq:Displacement})).
    Top panel shows the results  obtained with the initial state $ | \Psi_\text{qumode}^\text{ini}  \rangle = \ket{0}$ while below we have $| \Psi_\text{qumode}^\text{ini} \rangle  = \ket{3}$.
    It is noteworthy that, while the qumode space is theoretically infinite, we here represent evolution of the wavefunction within a restricted space by truncating the maximum number of bosons to 8.
    }
    \label{fig:CD_circuit}
\end{figure}

 \vspace{0.25cm}

\paragraph{{Hybrid Qubit-Qumode-Based Architectures.}} 

On this platform, we will take advantage of the digital-analog point of view offered by the hardware to encode electron-photon entanglement using a specific hybrid qumode-qubit gate called the \textit{Controlled-Displacement} gate~\cite{crane2024hybrid,liu2024hybrid}:  
\begin{equation}\label{eq:CD_gate}
    CD(\theta) = \exp\left( \hat{Z}_q \otimes \left( \theta \hat{b}  - \theta^* \hat{b}^\dagger \right) \right),
\end{equation}
where \(\theta \in \mathbb{C}\) is a complex number and $q$ is the control qubit.
Note here that we directly assimilated the qumode creation/annihilation operators to the photonic ones by posing the shorthand notation $b^\dagger_\text{qumode} = b^\dagger $.
As illustrated in Fig.~\ref{fig:CD_circuit} (and discussed in Refs.~\cite{crane2024hybrid,eickbusch2022fast}), based on hardware considerations, this hybrid qubit-qumode gate can be decomposed into three native operations, including two primitive entangling \textit{controlled-parity} gates  
\begin{equation}
    CR( \pi ) = \exp\left( -i \frac{\pi}{2} \hat{Z}_q \otimes \hat{b}^\dagger \hat{b} \right),
\end{equation}  
 sandwiching a pure bosonic \textit{Displacement}  gate,  
\begin{equation}\label{eq:Displacement}
    D(\theta) = \exp\left( \theta \hat{b} - \theta^* \hat{b}^\dagger \right).
\end{equation}  
such that we have
\begin{equation} \label{eq:parity}
    CD(\theta) = CR( \pi ) D(\theta) CR( \pi )^\dagger.
\end{equation}
This decomposition can be proven based on the Baker-Campbell-Hausdorff formula and using the property $CR(\pi) \hat{b} \, CR(\pi)^\dagger = i \hat{Z}_q \otimes \hat{b}$ (see Ref.~\cite{crane2024hybrid} for more details).

\begin{table*}[t]
\includegraphics[width=16.5cm]{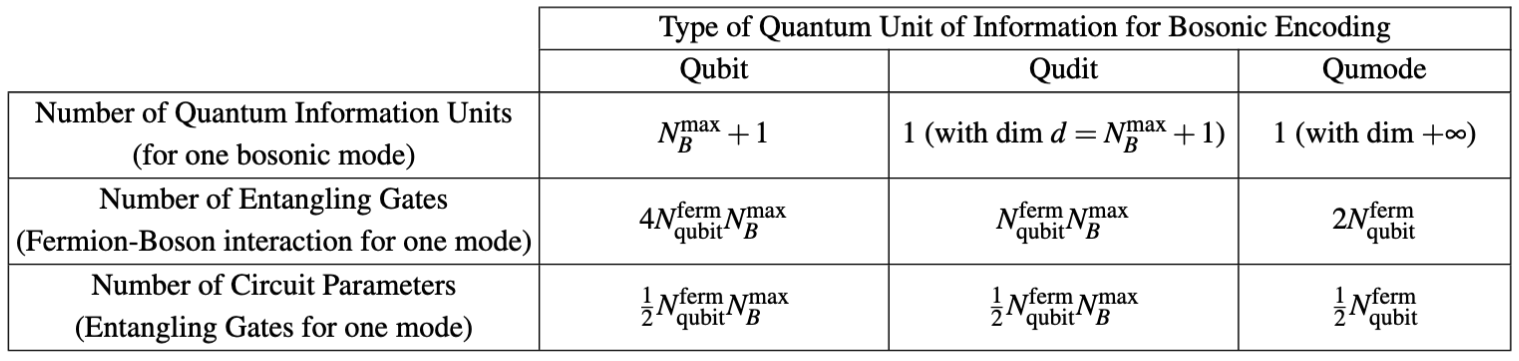}
\caption{
\textbf{Scaling of quantum resources required in our circuit designs adapted to qubit-, qudit- and qumode-based platforms.}\linebreak
We report the number of quantum information units required to encode bosonic degrees of freedom across the three platforms. 
Additionally, we present the entangling gate counts necessary to represent electron-photon entanglement within a single block of the proposed ansatz, along with the corresponding number of circuit parameters to be optimized. 
Here, $N_{\text{qubit}}^{\text{ferm}}$ denotes the number of qubit-like information units used on each platform to encode fermionic degrees of freedom, while $N_B^{\max}$ represents the maximum number of bosons considered in the cavity (truncation imposed for discrete-variables-based platforms, \textit{i.e.} the qubits and qudits ones).
}
\label{table:Counting_resources}

\end{table*}

In practice, the controlled-displacement gate $CD(\theta)$ induces a non-trivial transformation in the infinite qumode Fock basis, which is deeply rooted in the concept of coherent states (also known as Glauber states~\cite{glauber1963coherent,cahill1969ordered}).  
To better understand this effect, we observe from Eq.~(\ref{eq:CD_gate}) that the operation corresponds to a displacement operator whose phase is controlled by the expectation value of the $Z_q$ operator associated with the master control qubit $q$.  
Focusing then solely on the action of the displacement operator (see Eq.~(\ref{eq:Displacement})), we illustrate in the lower panel of Figure~\ref{fig:CD_circuit} the effect of the gate on two initial qumode states that contain either zero  $| \Psi_\text{qumode}^\text{ini}  \rangle = \ket{0} $, or three bosons $| \Psi_\text{qumode}^\text{ini}  \rangle = \ket{3}$.
As shown here, increasing the real part of the gate parameter $\theta$ leads to the propagation of an initial state throughout the qumode Fock space that covers many Fock states simultaneously. 
This large propagation arises because the displacement operator $D(\theta)$ actually generates a (non-linearly) parametrized quantum superposition of bosonic coherent states that will effectively span a large amount of the qumode Fock basis.  
This contrasts strongly with the effect of the controled-Givens operations introduced before for the two \textit{discrete-variable} platforms involving \textit{qubits} and \textit{qudits}.
 While Givens rotations $G(\theta)/\mathcal G(\theta)$ are ``local'' operations that couple two states at the same time within a finite truncated Hilbert space (see Eq.~(\ref{eq:givens_rotation}) and (\ref{eq:qudit_givens})),  the displacement gate $D(\theta)$ has a more extended action that can theoretically cover a large part of an infinite Hilbert space. 
 This will prove to be very efficient in the way to encode electron-photon entanglement, as we will show later on in this work.
Last but not least, note that while the $CD(\theta)$ gate is parametrized by a complex amplitude $\theta \in \mathbb{C}$, in this work we focus exclusively on its real parametrization, fixing $\Im( \theta ) = 0$.

\subsection{Quantum Circuits Ansatze:\\ Accounting for Quantum Resources on Each Platform}
 
We conclude this section with a detailed comparison of the quantum resources required across different platforms for implementing the circuit ans\"{a}tze that will describe the low-lying states of the polaritonic problem. 
Our analysis focuses on a single layer of the circuit as shown in Fig.~\ref{fig:Compact_circuits}, with particular attention given to the number of quantum units of information involved and the entangling operations between light and matter. 
Table~\ref{table:Counting_resources} serves as a reference guide for the numerical estimates presented in the following.

We begin by examining the number of quantum information units needed to encode a single photonic mode on each platform, as shown in the first row of Table~\ref{table:Counting_resources}. 
In \textit{discrete-variable} platforms, including both \textit{qubit} and \textit{qudit} platforms, the representable photonic subspace is truncated at a maximum of $N_B^{\text{max}}$ photons in the cavity. 
The qubit-based encoding then requires $N_B^{\text{max}} + 1$ units using a direct boson mapping~\cite{somma2002simulating, veis2016quantum, peng2023quantum, knapik2025quantum}.
However, the same photonic subspace can be more compactly represented using a single qudit of dimension $d = N_B^{\text{max}} + 1$, \textit{via} the mapping we introduced in Eq.~(\ref{eq:DBM_qudit}) based on generalized Gell-Mann matrices. 
In contrast, the \textit{continuous-variable} encoding employs a single \textit{qumode}, which can, in principle, represent the full photonic state space without truncation, owing to the infinite dimensionality of the mode.

Let us now focus on the number of entangling gates required to capture the light-matter correlations in each circuit. 
As illustrated in Fig.~\ref{fig:Compact_circuits}, for both the qubit and qudit implementations we adopt a specific ``stair-like'' architecture that connects each spin orbital to the $N_B^\text{max}$ nearest photonic transitions (e.g., $0 \leftrightarrow 1$ photon, $1 \leftrightarrow 2$ photons, etc.). 
Within this architecture, the number of controlled-Givens operations per layer scales as $N_{\text{qubit}}^{\text{ferm}} N_B^{\text{max}}$, where $N_{\text{qubit}}^{\text{ferm}}$ denotes the number of qubit-like information units used to encode the fermionic degrees of freedom (the number of spin orbitals in the molecular system, e.g. $N_{\text{qubit}}^{\text{ferm}} = 4$ for H$_2$ in a minimal basis). \linebreak
For qudit-based platforms, this expression directly yields the number of entangling gates, since controlled-Givens rotations $C\mathcal{G}(\theta)$ (defined in Eq.~(\ref{eq:Control_givens_qudit})) are assumed to be native entangling operations~\cite{fischer2023universal, omanakuttan2023qudit}. 
In contrast, on qubit-based hardware, each controlled-Givens gate $C G(\theta)$ (defined in Eq.~(\ref{eq:CGiven})) must be decomposed into four standard entangling operations, namely: two controlled-$X$ and two controlled-$R_Y$ gates (see Fig.~\ref{fig:Givens_circuits}). 
This results in a total of $4 N_{\text{qubit}}^{\text{ferm}} N_B^{\text{max}}$ entangling gates per layer for the qubit platform.
Note that this scaling reflects the chosen stair-like design, which provides a minimal and efficient structure to enable all-to-all coupling between electronic orbitals and photonic excitations. 
Turning now to the qumode-based platform, as shown in the lower right panel of Fig.~\ref{fig:Compact_circuits}, we observe that circuit architectures can become even more resource-efficient. 
In this case, each circuit layer contains $N_{\text{qubit}}^{\text{ferm}}$ control-displacement gates $CD(\theta)$ (defined in Eq.~(\ref{eq:CD_gate})).
As shown in Fig.~\ref{fig:CD_circuit}, each of such gate involves two primitive parity operations~\cite{crane2024hybrid,liu2024hybrid}, leading to a total of $2 N_{\text{qubit}}^{\text{ferm}}$ entangling gates per layer.

We now consider the number of variational parameters to be optimized in a VQE routine. For both the qubit and qudit platforms, it is equal to $\tfrac{1}{2} N_{\text{qubit}}^{\text{ferm}} N_B^{\text{max}}$ parameters per layer. \linebreak
This direct correspondence stems from the fact that, despite differences in hardware, the underlying circuit structures are mathematically equivalent, since both rely on the same actions of local Givens rotations. 
In contrast, qumode-based encoding requires only $\tfrac{1}{2} N_{\text{qubit}}^{\text{ferm}}$ parameters per layer. 
This considerable reduction is due to the expressivity of control-Displacement operations, which can generate states spanning the entire Fock space of the mode (in theory infinite-dimensional).

In summary, this resource comparison already highlights the potential interest in employing qudit and qumode quantum encodings over the conventional qubit-based approach. 
These platforms offer substantial reductions in terms of number of units of information, circuit depth and parameter count, making them especially promising alternative to encode fermion-boson problem on quantum devices with limited resources.
In the following section, we will illustrate the performance of these different approaches with numerical simulations.

\section{ Numerical Simulations and Discussion }
\label{sec:IV}

We will now benchmark our quantum circuit designs, adapted to qubit, qudit and hybrid qubit-qumode platforms, within the SA-VQE framework to describe the three low-lying states of the H$_2$ molecule in a cavity using state-vector simulations. 
To proceed, we first briefly summarize the numerical tools employed to emulate each platform to facilitate the reproducibility of our results.
We then present a comparative study that focuses on the number of circuit layers required across different light-matter coupling regimes. 
Finally, we conclude with an illustration of the capability of our approach to capture LIAC and LICI.

\subsection{ Computational Tools }  
\label{sec:CompDet}

To carry out our numerical study, we employ different computational packages depending on the target platform. 
Specifically, we use the \textit{OpenFermion} and \textit{Cirq} libraries~\cite{mcclean2020openfermion} to simulate qubit- and qudit-based architectures.
Notably, for qudit-based simulations, we extend the existing functionality in \textit{OpenFermion} to support higher-dimensional quantum states, including the implementation of GGM and their associated gates (see Ref.~\cite{Qudits_QC}).
For hybrid qubit-qumode simulations, we utilize the recently developed \textit{Bosonic-Qiskit} package~\cite{stavenger2022c2qa}.
As a reference system, we consider the hydrogen molecule H$_2$, described in a minimal STO-3G basis, embedded in an optical cavity containing a single reference photonic mode. The coupling strength and mode frequency are specified in subsequent sections.
All reference QED-FCI calculations are performed using the open-source package \textit{QuantNBody}~\cite{yalouz2022quantnbody}, which enables the systematic construction of second-quantized operators and hybrid fermion-boson Hamiltonians, including cases where boson number conservation is broken. 
This is combined with the quantum chemistry software \textit{Psi4}~\cite{smith2020psi4}, which provides the electronic integrals necessary to simulate the Pauli-Fierz Hamiltonian (see Eq.~(\ref{eq:Pauli_fierz_ham})).
Quantum circuit parameters are optimized using the Sequential Least Squares Programming (SLSQP) algorithm from the \textit{SciPy} library.
For the SA-VQE procedure, we consider the following initial light-matter configurations: \textit{(i)} a Hartree–Fock state with zero photon, $\ket{\psi_A} = \ket{\text{HF}} \otimes \ket{0}$; \textit{(ii)} a HOMO-LUMO singlet-excited Hartree–Fock state with zero photon, $\ket{\psi_B} = \tfrac{1}{\sqrt{2}} \hat{E}_{\text{H-L}} \ket{\text{HF}} \otimes \ket{0}$; and \textit{(iii)} a Hartree-Fock state with one photon, $\ket{\psi_C} = \ket{\text{HF}} \otimes \ket{1}$.
Finally, note that in all simulations involving discrete-variable platforms (\textit{i.e.} qubits and qudits), the photonic Fock space is truncated to a maximum of $N_B^{\text{max}} = 3$ photons. 
Regarding qumode emulation, although the qumode platform is theoretically infinite-dimensional, its simulation using \textit{Bosonic-Qiskit} necessarily requires a finite Fock space representation (see Ref.~\cite{stavenger2022c2qa}). \linebreak
To mitigate potential artifacts arising from a ``hard'' truncation and to more accurately approximate the true qumode Fock space, we adopt a larger cutoff of a maximum number of 15 photons in our qumode-based simulations.
In the different studies presented in the following, we try to inspect different situations and light-matter coupling regimes, which still guarantee to provide a fair comparison between the different platforms considered. 
For the sake of conciseness, we do not delve into details here, but we refer interested readers to the Appendix~\ref{app:Fair Comp}, where these points are discussed more thoroughly. 


\subsection{ SA-VQE Energies at Cavity Resonance:\\ Accuracy with Circuit Depth   }

\begin{figure}[t]
    \centering
    \includegraphics[width=8cm]{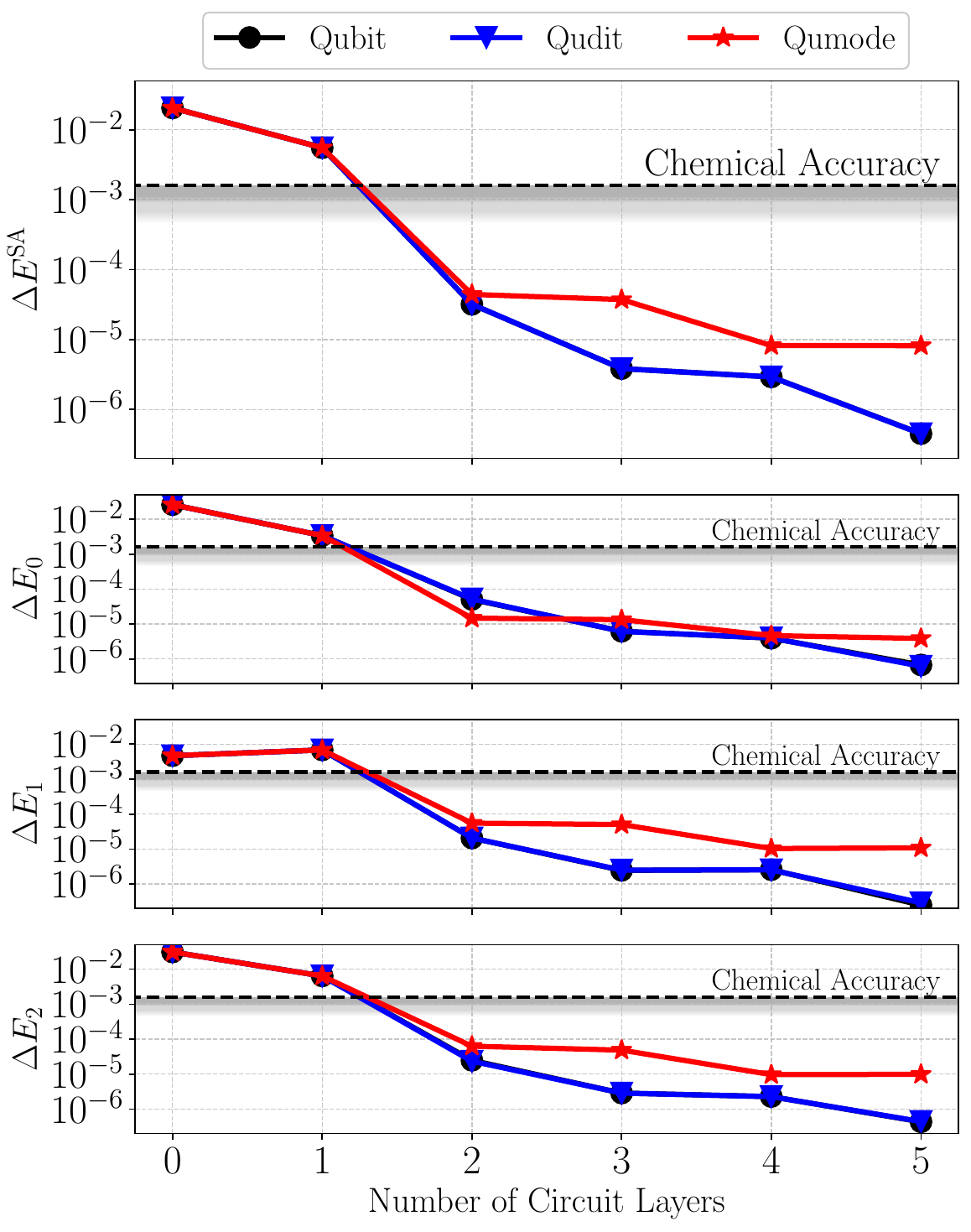}
    \caption{\textbf{Evolution of the energy errors obtained with SA-VQE as a function of the number of circuit layers.}
     All results are obtained for the H$_2$ molecule at geometry $r = 0.74$ \ \AA\  with cavity frequency $\omega = 1$ Ha and coupling $\lambda = 0.05$. 
     Results corresponding to the  qubit, qudit, and qumode platforms are depicted using black curves with circles, blue curves with triangles and red curves with stars, respectively.
    \textbf{Top panel:} Evolution of the SA-VQE ensemble energy error $\Delta E^\text{SA}$ as given in Eq.~(\ref{eq:SA_accuracy}).
    \textbf{Three lower panels:} Evolution of the SA-VQE state specific energy error $\Delta E_k$ as given in Eq.~(\ref{eq:SS_Energy_error}) for the ground, first and second excited states.}
    \label{fig:Energy-accuracy}
\end{figure}

In the first part of our study, we illustrate how the different quantum circuit ansätze introduced in Sec.~\ref{sec:CompactAnsatz} perform in the description of the energies of the three low lying polaritonic states of H$_2$ in cavity.
To this end, we perform a single point calculation at the interatomic distance $r = 0.74$~\AA, corresponding to the molecular geometry where resonance with the cavity mode ($\omega = 1 $ Ha) induces the formation of a  LIAC (\textit{i.e.} Rabi splitting) between the first and second polaritonic excited states, as previously illustrated in Fig.~\ref{fig:H2_illustration}.
This case study therefore represents the scenario in which light-matter coupling produces the richest polaritonic signatures.
In this study, we considered a fixed light-matter coupling $\lambda = 0.05$ that corresponds to a weak/intermediate interaction regime.
To evaluate the performance of the ans\"{a}tze within SA-VQE, we analyzed how the resulting energy accuracy evolves as the depth of the quantum circuit is progressively increased (\textit{i.e.} the number of layers). 
Figure~\ref{fig:Energy-accuracy} illustrates the evolution of the energy error compared to the QED-FCI results for each platform.
The results corresponding to the qubit, qudit and qumode platforms are depicted using black curves with circles, blue curves with triangles, and red curves with stars, respectively.

Let us first focus on the top panel of Figure~\ref{fig:Energy-accuracy} which shows the absolute error of the ensemble energy $\Delta E^{\rm SA}$ defined as
\begin{equation}\label{eq:SA_accuracy}
\Delta E^\text{SA} = \abs{ E^\text{SA-VQE}(\vec{\theta}^\star) - E^\text{QED-FCI}_\text{Ens}},
\end{equation} 
which quantifies the deviation between the optimized SA-VQE energy, as defined in Eq.~(\ref{eq:SA-VQE-energy}) (with $\vec{\theta}^\star$ the optimal circuit parameters) and the exact ensemble QED-FCI energy
\begin{equation}
    E^\text{QED-FCI}_\text{Ens} = \frac{1}{3} \sum_{k=0}^2 E^\text{QED-FCI}_k.
\end{equation} 
As clearly observed in the upper panel of Fig.~\ref{fig:Energy-accuracy}, starting from the initial unparametrized states (\textit{i.e.} with zero circuit layers), the ensemble energy error is already quite low with $ \Delta E^\text{SA} \sim 10^{-2}$ Ha. 
This indicates that the initial states we considered (see previous section) already provide a reasonable approximate of the three exact low-lying polaritonic eigenstates.
For all platforms considered, increasing the number of circuit layers systematically reduces the error of the ensemble energy $\Delta E^\text{SA}$.
Remarkably, with only two layers, the three platforms reach convergence within chemical accuracy, defined as $\Delta E^\text{SA} <$ 1.6 mHa (limit shown in Fig.~\ref{fig:Energy-accuracy} by a horizontal black dashed line).
Further increasing the circuit depth continues to lower the error, eventually reaching values below $10^{-6}$~Ha for qubit/qudits platforms and $10^{-5}$~Ha for the qumode one.
The results clearly show that higher ensemble energy accuracy can, in principle, be achieved by using deeper circuit ansätze, owing to their increased expressibility.
Interestingly, a similar trend is observed when focusing on the state-specific energy errors for the three SA-VQE polaritonic states defined as
\begin{equation}\label{eq:SS_Energy_error}
\Delta E_k = \abs{ E_k^\text{SA-VQE}(\vec{\theta}^\star) - E_k^\text{QED-FCI} },
\end{equation}
where the index $k = 0$, 1, and 2 correspond to the ground state and the first and second polaritonic excited states, respectively.
Note here that the $E_k^\text{SA-VQE}(\vec{\theta}^\star)$ are obtained after a final state resolution of the SA-VQE procedure involving the classical diagonalization of a 3x3 Hamiltonian matrix as explained in Sec.~\ref{sec:SAVQE}. 
The results of the state specific errors $\Delta E_k$  are presented in the three lower panels of Fig.~\ref{fig:Energy-accuracy}.
In the latter panels, we clearly observe that each individual state energy converges toward a more accurate approximation of the exact QED-FCI reference as the number of layers increases.
Remarkably, the dispersion of the individual energy errors remains steady across different circuit depths, indicating that all three state energies are consistently well-approximated, with no specific state being disproportionately favored or neglected.

To conclude this first section, we want to highlight two important points.
First, all panels in Figure~\ref{fig:Energy-accuracy} show that the three platforms yield broadly similar results in terms of energy accuracy. 
This consistency supports the reliability of the different ansatz circuits introduced, regardless of the underlying hardware architecture.
Second, and more specifically, the qubit and qudit platforms display nearly identical trends. 
This similarity is expected, as both platforms, despite differences in their physical implementation, rely on the same type of mathematical transformation, namely: \textit{Givens rotations} (see Sec.~\ref{sec:Entangling gates}).
The minor discrepancies observed at greater circuit depths and in regimes of very small energy errors can be attributed to differences in the numerical emulation of the two platforms, which use matrix representations of different dimensions.

\begin{figure*}
    \centering
    \includegraphics[width=16cm]{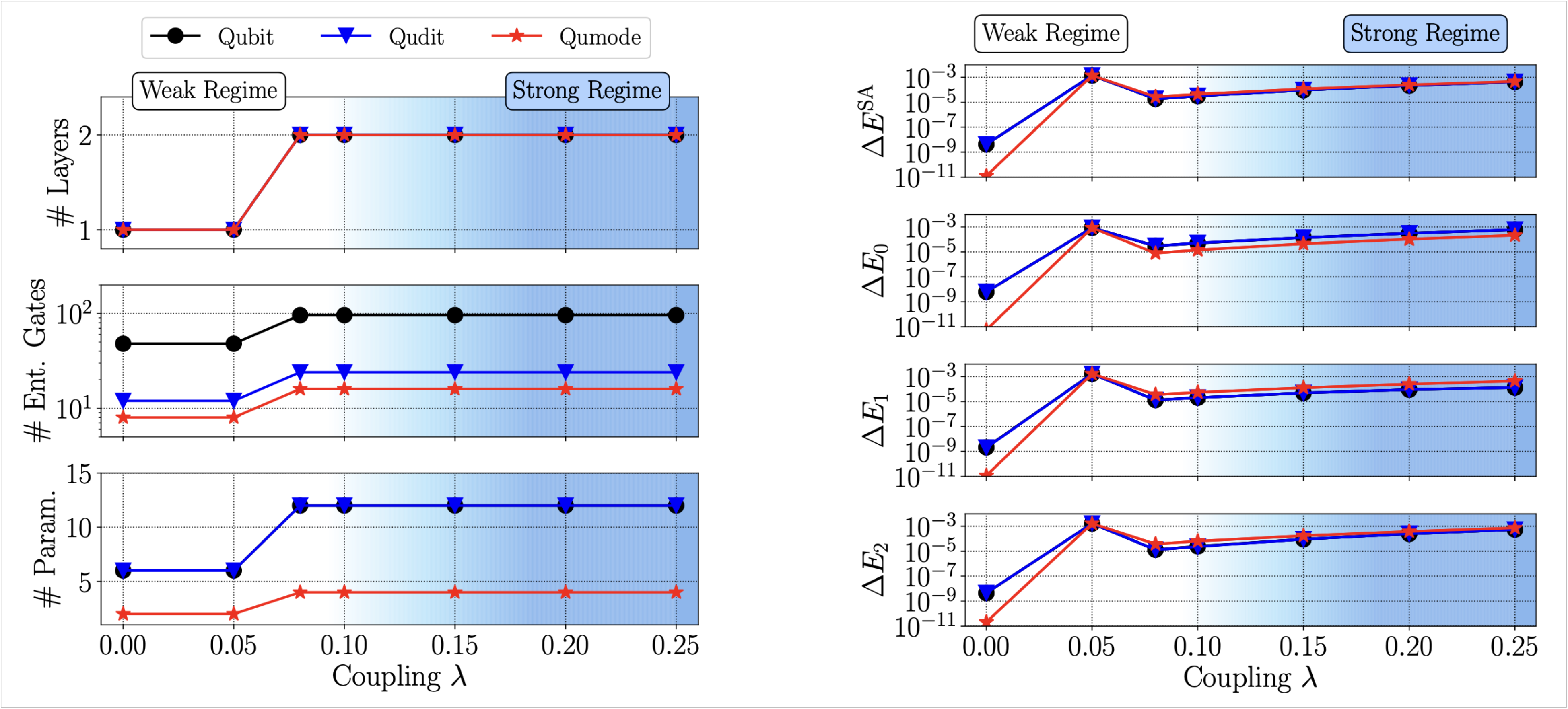}
    \caption{\textbf{Evolution of the circuit complexity as a function of the coupling parameter $\lambda$ for the three platforms considered.} 
   All results are obtained using the SA-VQE algorithm applied to the H$_2$ molecule in cavity at geometry $r = 0.74$ \ \AA, where resonance with the cavity is observed with $\omega = 1$ Ha. 
   Circuit complexity is quantified by the number of layers required to reach chemical accuracy (\textit{i.e.} a threshold of 1.6 mHa) on the ensemble energy of the three low-lying states of the system.
   \textbf{Left part:} 
Top panel show the number of circuit layers required for each platform to achieve chemical accuracy as $\lambda$ increases.
Middle panel and bottom panel show the associated number of entangling gates necessary to generate the light-matter entanglement.
   \textbf{Right part:} Resulting energy errors compared to QED-FCI for the three platforms.
   Top to bottom panels respectively show energy errors for the ensemble, ground, first- and second-excited state as given in Eqs.~(\ref{eq:SA_accuracy},\ref{eq:SS_Energy_error}).
    }
    \label{fig:Layers_count}
\end{figure*}

\subsection{ Assessing  Quantum Circuits Complexity \\ To Describe Different Light-Matter Coupling Regimes}


In this second part of our study, we wanted to assess the performance of the different quantum circuit ansätze in describing the three low-lying polaritonic states of H$_2$, but now, as the light-matter coupling strength $\lambda$ varies (spanning regimes from weak to strong interaction).
To proceed, we decided to focus on a practical question: for a given value of $\lambda$, \linebreak how many circuit layers would be needed to achieve chemical accuracy on the ensemble energy (i.e., $\Delta E^\text{SA} <$ 1.6 mHa)?
Beyond probing the effects of light-matter interaction on the complexity of the polaritonic states, by addressing this question, we also gain valuable insights regarding the minimal number of quantum resources required on each platform to meet a practical computational objective. 
This makes important sense in the NISQ era, where hardware limitations are present and where, in principle, deep circuits are prohibitive.
To carry out our study, we considered the interatomic distance $r = 0.74 $ \AA\ for which resonance with the cavity frequency $\omega = 1$ Ha leads to the emergence of a LIAC (see Fig.~\ref{fig:H2_illustration}). 
We then systematically varied the coupling amplitude in the range $\lambda \in [0, 0.25]$, and determined the minimum number of circuit layers required on each platform to generate an ensemble energy with a relative error $\Delta E^\text{SA} <  $ 1.6 mHa.

 



In the left column of Figure~\ref{fig:Layers_count}, we show the evolution of the number of circuit layers required for each of the three platforms as a function of $\lambda$.  
The results corresponding to the qubit, qudit and qumode platforms are depicted using black curves with circles, blue curves with triangles, and red curves with stars, respectively.
As readily seen here, regardless of the platform, we observe that only a few layers of the different circuit ans\"{a}tze are sufficient to reach chemical accuracy regardless of the coupling regime considered.
In the weak-coupling regime, where $\lambda \leq 0.05$, a single layer is sufficient to describe the ensemble energy within the target chemical accuracy.
However, this changes when we enter the stronger coupling regime. 
Indeed, as $\lambda > 0.05$ and increases, we observe in the top panel of Fig.~\ref{fig:Layers_count} a corresponding increase in the number of layers required, reaching a plateau value of two layers that remains steady across all the coupling values explored up to the strong limit value of $\lambda = 0.25$.
Interpretations regarding this increase will be provided in the next paragraphs.

Let us now focus on the estimate of quantum resources required on each platform to produce these results within chemical accuracy.
In the middle panel of Fig.~\ref{fig:Layers_count} (left column) we illustrate the number of light-matter entangling gates necessary for each platform (following the rules presented in Table~\ref{table:Counting_resources}).
Here, we see that while the three architectures yield similar results in terms of layer counts, the number of entangling gates differs significantly.
Across all regimes explored, we find that the qumode-based platform is the least resource intensive, requiring at most 16 entangling gates, followed by the qudit platform with 24 gates, and the qubit platform with 96 gates.\linebreak
In the bottom panel just below (left column), we also show the corresponding number of circuit parameters to be optimized within the SA-VQE algorithm. 
Here again, the result highlights the advantage of the hybrid qubit-qumode platform, which requires (at most) 4 parameters in total compared to the qubit- and qudit-based platforms, which both require 16 circuit parameters.
It is worth noting that the results presented here are quite general, as we conducted similar studies at different geometries (not shown here) and observed very similar behavior in terms of the relative resource requirements across platforms.

\begin{figure*} 
    \centering
    \includegraphics[width=18cm]{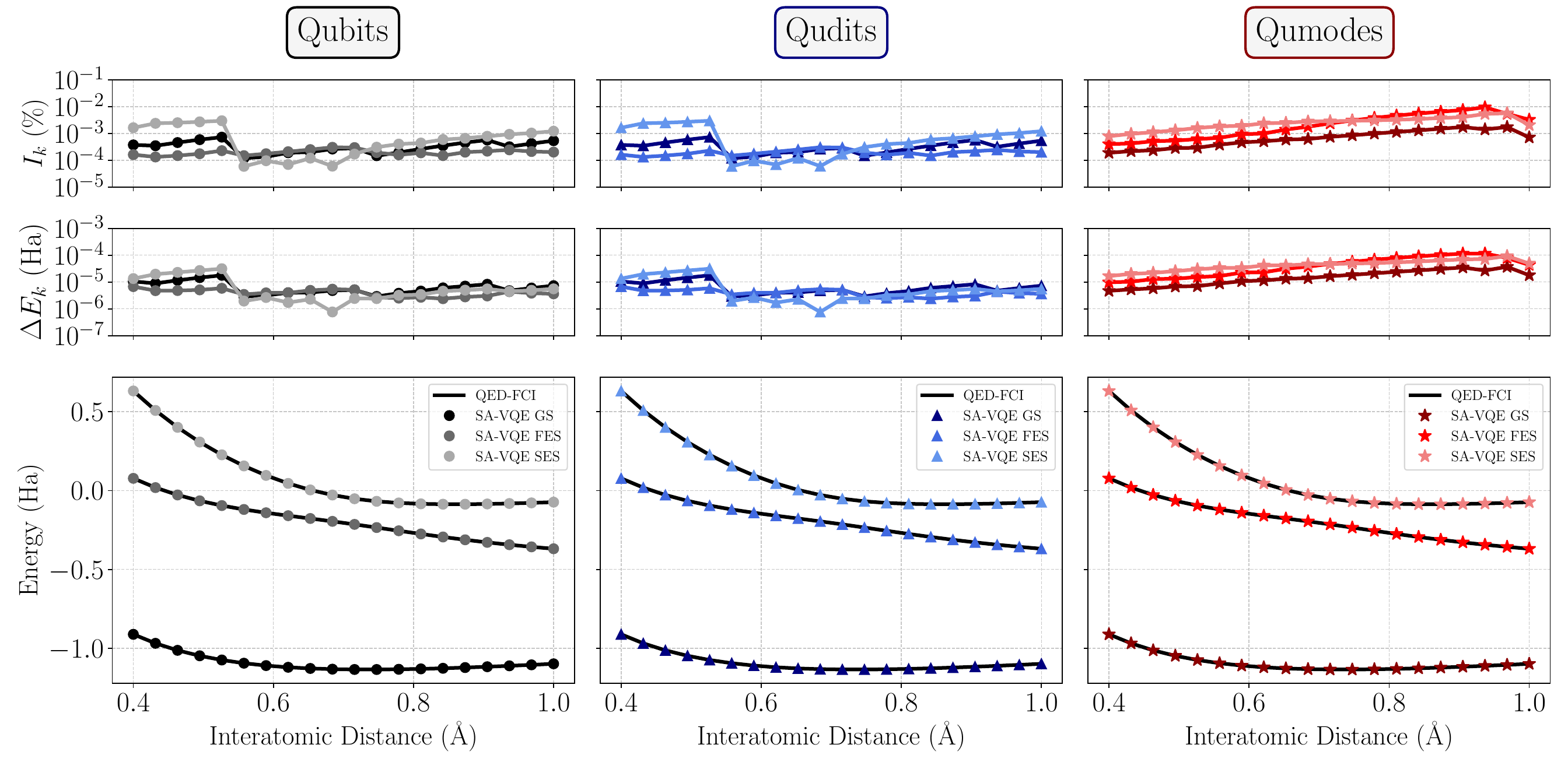}   
\caption{  \textbf{SA-VQE Potential Energy Surfaces for a Light-Induced Avoided Crossing Across Qubit, Qudit, and Qumode Platforms.} \linebreak
The left, middle, and right columns respectively present SA-VQE results obtained using qubit, qudit, and qumode based platforms. All results are compared against equivalent reference QED-FCI calculations for a cavity frequency $\omega = 1$ Ha and a coupling $\lambda = 0.1$. 
\textbf{Top row:} Infidelity $I_k$ of the three lowest SA-VQE polaritonic states, as defined in Eq.~(\ref{eq:fidelity}). 
\textbf{Middle row:} State specific absolute energy errors $ \Delta E_k$ as defined in Eq.~(\ref{eq:SS_Energy_error}). 
\textbf{Bottom row:} Energy spectra comparing QED-FCI energies (solid black lines) with SA-VQE estimates (colored markers). 
}
    \label{fig:PES}
\end{figure*}

To complete our analysis, the right column of Figure~\ref{fig:Layers_count} shows the evolution of the energy errors obtained for each value of the light–matter coupling $\lambda$ (evaluated at the corresponding number of circuit layers required to reach the chemical accuracy as shown in the left column of Figure~\ref{fig:Layers_count}). 
From top to bottom, the panels display the ensemble energy error $\Delta E^{\text{SA}}$, followed by the individual state-specific errors $\Delta E_0$, $\Delta E_1$, and $\Delta E_2$, as defined in Eqs.~(\ref{eq:SA_accuracy}) and (\ref{eq:SS_Energy_error}), respectively.
As can be readily observed, at zero coupling $\lambda = 0$, where only one circuit layer is required, all energy errors are extremely small across all platforms. 
However, as the coupling strength increases $\lambda \geq 0.05$ and enters stronger interaction regimes (requiring two circuit layers), the errors suddenly jump and then gradually increase with $\lambda$. 
To understand this behavior, numerical simulations focused on the structure of the exact QED-FCI states have been carried out. 
For conciseness, we will not show these results here, but we briefly refer to our observations (for more details, we direct interested readers to Appendix~\ref{app:WF_structure} where these state decompositions are illustrated). 
The main results of our simulations reveal that as $\lambda$ increases, the many-body nature of the three low-lying polaritonic eigenstates becomes progressively more complex, involving contributions not only from the 0- and 1-photon subspaces but also non-negligible components from the 2- and 3-photon subspaces. 
This increasing complexity necessitates more expressive circuits for accurate representation, specifically as here we start from three initial states which do not even present any contribution on these subspaces (as detailed in previous Sec.~\ref{sec:CompDet}).
This explains both the increase in the number of required circuit layers (left column of Figure~\ref{fig:Layers_count}) and the continued growth of the energy error even after increasing the circuit expressivity (right column of Figure~\ref{fig:Layers_count}) due to the increasing complexity of the states to be encoded in a circuit.

\begin{figure*} 
    \centering
    \includegraphics[width=17cm]{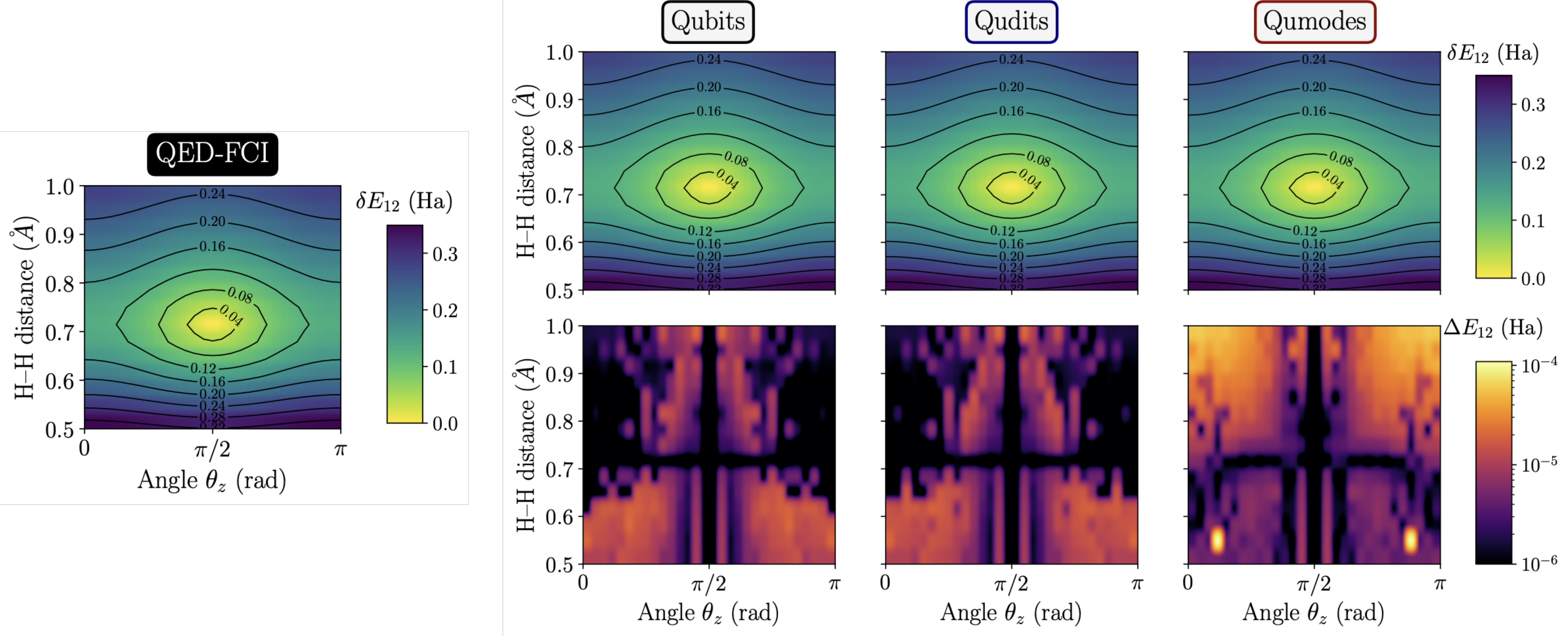} 
    \caption{ 
\textbf{ SA-VQE results for the description of the Light-Induced Conical Intersection for the H$_2$ molecule in cavity.}  
 Here we represent the vertical excitation energy between the first and second polaritonic eigenstates in the ($r,\theta_z$)-parameter space for H$_2$ with coupling strength $\lambda = 0.08$ and photon frequency $\omega = 1$ Ha.  
\textbf{Left part:}  heatmap showing the reference QED-FCI results where the yellow central zone indicate the location of the LICI.
\textbf{Right part:} results obtained with the qubit, qudit and qumode platforms.
First row show heatmaps of the vertical excitation energy while second row show the associated energy error compared to QED-FCI as given in Eq.~(\ref{eq:Vert_Energy_error}).
As readily seen here, all the three platforms present results that are in strong agreement with the QED-FCI reference throughout the whole ($r,\theta_z$)-parameter space.
}
    \label{fig:3D_LICI_SA_VQE}
\end{figure*}

To conclude here, we want to emphasize again an important point:
remarkably, all the results presented so far reveal that the evolution of energy errors across the three platforms remains consistent regardless of the conditions considered.
This reinforces the idea that, despite relying on distinct hardware implementations and varying amounts of quantum resources (see Fig.~\ref{fig:Layers_count}), the different ansatz circuits proposed for the qubit-, qudit- and qumode-based platforms are all reliable, achieving qualitatively similar levels of accuracy in a wealth of contexts.

\subsection{  SA-VQE Polaritonic PES: \\ 
Light Induced Avoided Crossing and Conical Intersection}

In this third and last part of our study, we will provide numerical proofs highlighting the performance of the three quantum circuit ansatze in describing the hallmark polaritonic spectral features, namely the LIAC and LICI of H$_2$ in cavity.
In all the following simulations, we will consider the same fixed number of 3 circuit layers regardless of the platform.

In the first part of our study, we will focus on the generation of a one-dimensional PES of the LIAC that arises between the first and second polaritonic excited states in the region around $r = 0.74$~\AA\ (resonance with cavity mode frequency $\omega = 1$ Ha).\linebreak
We fix the light-matter coupling strength to $\lambda = 0.1$ (medium/strong regime) and scan the interatomic distance in the range $r = [0.4\text{\AA}, 1 \text{\AA}]$, which corresponds to the relevant geometries for observing the LIAC structure. 
Figure~\ref{fig:PES} presents, in three columns from left to right, the results obtained for the qubit-, qudit-, and qumode-based platforms, respectively.
The first row shows the state-specific infidelity, denoted $I_k$ and defined by 
\begin{equation}\label{eq:fidelity}
I_k = 1 -  | \langle \Psi^{\text{ QED-FCI}}_k | \Psi^{\text{ SA-VQE}}_k \rangle  |^2,
\end{equation}
which quantifies the similarity between SA-VQE states and QED-FCI eigenstates ($I_k = 0$, which implies a perfect similarity).
The second and third rows in Figure~\ref{fig:PES} show the state-specific energy errors (as given in Eq.~(\ref{eq:SS_Energy_error})) and the PES of the three low-lying polaritonic states, respectively.
Within each column, we use shades of a single color and the same family of markers to depict the results associated with different SA-VQE states (gray circles for qubits, blue triangles for qudits, and red stars for qumodes), while QED-FCI reference energies are shown with solid black lines (last row).
As readily seen in Fig.~\ref{fig:PES},  the characteristic shape of the Rabi splitting is very well captured with associated features. 
Across all scanned PES, the state-specific energy errors are consistently below $10^{-4}$~Ha, and the state infidelities are always lower than $0.01$ \%.
All these results reveal that the SA-VQE states are an excellent approximation of the exact QED-FCI reference $| \Psi^{\text{\tiny SA-VQE}}_k \rangle \sim | \Psi^{\text{\tiny QED-FCI}}_k \rangle $.
Once again, the simulations produced for each hardware platform exhibit very similar trends and accuracy (for a same fixed number of circuit layer).

To conclude our numerical investigations, we sought to evaluate whether similar behavior could be observed when considering the full characterization of the LICI for H$_2$ in cavity, accounting for both the interatomic distance and the rotation angle $\theta_z$ (as initially illustrated in Fig.~\ref{fig:H2_illustration} for $\lambda = 0.08$ and $\omega = 1$ Ha).
Figure~\ref{fig:3D_LICI_SA_VQE} displays the results of our numerical simulations across the three hardware platforms. On the left, we present the reference QED-FCI vertical excitation energy defined by:
\begin{equation}
\delta E^\text{QED-FCI}_{12} = E_2^\text{QED-FCI} - E_1^\text{QED-FCI},
\end{equation}
which corresponds to the energy gap between the first and second polaritonic excited states and captures the key energetic feature of the LICI (typically the degenerated point $ E_1 = E_2$ is depicted by the central yellow central region).
The right panels of Fig.~\ref{fig:3D_LICI_SA_VQE} show the corresponding results obtained from SA-VQE simulations on each of the three platforms. The upper row reports the computed vertical excitation energies, while the lower row quantifies the error with respect to the QED-FCI reference.
This error, denoted $\Delta E_{12}$, is defined as:
\begin{equation}\label{eq:Vert_Energy_error}
\Delta E_{12} = \left| \delta E^\text{SA-VQE}_{12}(\vec{\theta}^\star) - \delta E^\text{QED-FCI}_{12} \right|,
\end{equation}
where $\delta E^\text{SA-VQE}_{12}(\vec{\theta}^\star) = E_2^\text{SA-VQE}(\vec{\theta}^\star) - E_1^\text{SA-VQE}(\vec{\theta}^\star)$ is the vertical excitation energy obtained with SA-VQE and $\delta E^\text{QED-FCI}_{12}$ is the exact reference value (as defined in Eq.~(\ref{eq:SS_Energy_error})).
As illustrated, all three platforms demonstrate both qualitative and quantitative agreement with the QED-FCI benchmark, with errors consistently remaining below $\Delta E_{12} < 10^{-4}$ Ha throughout the whole parameter space.
Additional simulations (not shown here) revealed that similar results are obtained for a larger maximum amplitude of coupling strength (\textit{i.e.} $\lambda = 0.25$).

\section{ Conclusion }
\label{sec:V}  

Motivated by the ambition to implement \textit{ab initio} polaritonic chemistry on emerging quantum computers, this prospective work focuses on a key question: which quantum computing platforms might, in the long term, be best suited to accurately describe and simulate hybrid light-matter systems? 
The complexity of this challenge lies in two key aspects, both of which are addressed in this work. 
First, we identified quantum architectures capable of accurately and efficiently handling the dual nature of polaritonic systems, which couple fermionic (electronic) and bosonic (photonic) degrees of freedom. 
Second, we developed concrete computational strategies to solve such problems on these platforms at an affordable cost, particularly within the constraints of the NISQ era.

To investigate this, we systematically explored and benchmarked three distinct computational paradigms for solving cavity quantum electrodynamics problems: \textit{(i)}~conventional qubit-based encodings, \textit{(ii)}~qudit-based approaches, and \textit{(iii)}~hybrid qubit-qumode architectures.  
For each platform, we developed compact, physically motivated quantum circuit ans\"{a}tze, integrating them within a state-averaged variational quantum eigensolver framework to simultaneously compute multiple polaritonic states.  
In designing these circuits, we incorporated entangling structures involving high-dimensional quantum gates, such as controlled-$\Lambda^Y$ gates and controlled displacement operations, which play a crucial role in our qudit-based and qumode-based strategies.
These gate types are inspired by recent developments, which have highlighted their potential realization on various quantum hardware (see Refs.~\cite{ fischer2023universal,omanakuttan2023qudit,brennen2005criteria,bullock2005asymptotically,muthukrishnan2000multivalued,crane2024hybrid,liu2024hybrid,eickbusch2022fast} and references within).

Our results, obtained for a prototypical H$_2$ molecule embedded in a cavity, allowed to highlight several important insights which we summarize here.

First, we demonstrated that all three platforms can accurately resolve the low-lying polaritonic spectrum of this system and reproduce hallmark light-matter phenomena, such as light-induced avoided crossings and conical intersections.  
These outcomes validate the physical expressivity of our tailored circuit ans\"{a}tze which leverage high dimensional quantum operations (for qudit- and qumode-based platforms).

Second, we emphasized a more critical point: although all platforms can, in principle, achieve similar levels of accuracy, they differ substantially in quantum resource requirements.  
Among them, the hybrid qubit-qumode approach offered the most favorable trade-off between the circuit depth, the number of information units, and the accuracy. 
This advantage stems from its ability to natively exploit the infinite-dimensional nature of bosonic modes to efficiently represent photonic excitations.  
The qudit-based strategy, while slightly more resource intensive, still showed significant gains over standard only-qubit-based methods, thanks to its access to a higher-dimensional local Hilbert space.  
In contrast, the conventional qubit-based approach required deeper circuits and more entangling operations, thereby limiting its scalability in near-term implementations, especially when treating bosonic subsystems.

Importantly, since all three hardware paradigms remain in active experimental development, this study is inherently prospective. 
Nevertheless, it opens several promising avenues for future research. 
First, it underscores the importance of adopting hardware-conscious encoding strategies that are tailored to the physical characteristics of the problem under consideration. 
In polaritonic systems, where hybrid fermion-boson correlations play a central role, platforms that natively support qudit or ``bosonic'' operations may offer compelling long-term advantages. 
Second, as hardware for qudits and qumodes continues to mature, the practical realization of these more efficient encodings becomes increasingly feasible. 
In this context, the compact entangling ans\"{a}tze introduced here could prove broadly useful in other domains of quantum chemistry and quantum optics.
As an opening, applying similar ans\"{a}tze to a broader range of hybrid systems, including larger molecules in cavities described with active spaces or electron phonon systems, could represent a compelling direction for future research and applications.

Ultimately, we believe that this work highlights the value of moving beyond conventional qubit-based paradigms (which often represent a rigid viewpoint within the quantum computing community) and embracing new quantum representations that align more naturally with the structure of the problem we want to address. 
Such an approach will be essential for unlocking the full potential of quantum computing in polaritonic chemistry and for advancing the simulation of complex light-matter interactions in the longer term.


\section*{Acknowledgement}
This work was supported by the Interdisciplinary Thematic Institute SysChem, through the IdEx Unistra (ANR-10-IDEX-0002), as part of the French Investments for the Future Program.
All authors acknowledge Prof. E. Fromager and Prof. V. Robert for the many insightful scientific discussions shared together.

\appendix

\section{Demonstration of commutation rules with spin operators $\hat{S}^2$ and $\hat{S}_z$ for the blocks of fermion-boson entangling gates employed in all circuit ansatze }
\label{app:proof_symmetry}

Assuming a quantum circuit that encodes both fermionic and bosonic degrees of freedom into distinct quantum registers, we analyze the properties of a parameterized controlled-$U$ gate, defined as
\begin{equation} \label{eq:app_controlled_U}
    C_q\hat{U}(\theta) = \ketbra{0_q}{0_q} \otimes \hat{\bold{1}} + \ketbra{1_q}{1_q} \otimes \hat{U}(\theta),
\end{equation}
where the control qubit $q$ resides in the first register, representing fermionic modes via the Jordan-Wigner mapping, while the unitary operator $\hat{U}(\theta)$ acts on the second register, which encodes bosonic modes (qubits, qudits, or qumodes). Specifically, we prove that applying two such gates consecutively, each controlled by a qubit associated with the spin-orbitals (up and down) of the same spatial orbital, ensures commutation with both the total spin operator $\hat{S}^2$ and the spin projection operator $\hat{S}_z$.

To proceed, we define the unitary operator $\hat{U}(\theta)$ acting solely on the bosonic register as
\begin{equation}
    \hat{U}(\theta) = \exp(\theta \hat{A}),
\end{equation}
where $\hat{A}$ is a generic anti-Hermitian operator satisfying $\hat{A}^\dagger = -\hat{A}$. Without loss of generality, note that $\hat{U}(\theta)$ can act on a single or multiple units of information within the bosonic register. 
Rewriting the controlled-$U$ gate in exponential form, we obtain
\begin{equation}\label{eq:control_U_qubit}
    C_q\hat{U}(\theta) = \exp\left( \theta \frac{1}{2} (\hat{\bold{1}}_q - \hat{Z}_q) \otimes \hat{A} \right).
\end{equation}
Under the Jordan-Wigner mapping (see Eq.~(\ref{eq:Jordan_Wigner})), we assume that the control qubit $q$ corresponds to a spin-orbital indexed by $(p,\sigma)$ with $p$ spatial orbital index and $\sigma \in  \lbrace \uparrow, \downarrow \rbrace$ spin index.
Within this framework, the controlled-$U$ operation can be rewritten as
\begin{equation}
    C_{p\sigma}\hat{U}(\theta) = \exp(\theta \hat{n}_{p\sigma} \otimes \hat{A}),
\end{equation}
where $\hat{n}_{p\sigma} = \hat{a}^\dagger_{p\sigma} \hat{a}_{p\sigma}$ is the spin-orbital number operator in second quantization.

Now, considering the product of two consecutive controlled-$U$ operations acting on the spin-orbitals (up and down) of the same spatial orbital $p$, we obtain
\begin{equation} \label{eq:exp}
\begin{split}
    C_{p\uparrow}\hat{U}(\theta_\uparrow) &C_{p\downarrow}\hat{U}(\theta_\downarrow) \\
    &= \exp(\theta_\uparrow \hat{n}_{p\uparrow} \otimes \hat{A} )
    \exp(\theta_\downarrow \hat{n}_{p\downarrow} \otimes \hat{A} ) \\
    &= \exp\left( \frac{1}{2} \left[ ( \theta_\uparrow+\theta_\downarrow) \hat{n}_p + (\theta_\uparrow-\theta_\downarrow )\hat{\zeta}_p \right] \otimes \hat{A} \right),
\end{split}
\end{equation}
where we have used the commutativity of the generators. Here, we introduce the spatial-orbital occupation number operator $\hat{n}_p = \hat{n}_{p\uparrow} + \hat{n}_{p\downarrow}$ and the spin polarization operator $\hat{\zeta}_p = \hat{n}_{p\uparrow} - \hat{n}_{p\downarrow}$.
As shown in Ref.~\cite{helgaker2013molecular}, while $\hat{n}_p$ always commutes with both $\hat{S}_z$ and $\hat{S}^2$, the spin-polarization operator only commute  with  $\hat{S}_z$  but not with $\hat{S}^2$ due to
\begin{equation}
\begin{split}
    \left[\hat{S}^2,\hat{\zeta}_p\right] &= 2\left[\hat{S}^2,\hat{n}_{p\uparrow}\right] = -2\left[\hat{S}^2,\hat{n}_{p\downarrow}\right] \\
    &= 2\left(\hat{S}_+\hat{a}_{p\downarrow}^\dagger\hat{a}_{p\uparrow} - \hat{a}_{p\uparrow}^\dagger\hat{a}_{p\downarrow}\hat{S}_-\right),
\end{split}
\end{equation}
where $\hat{S}_+ = \hat{S}_-^\dagger = \sum_p \hat{a}^\dagger_{p,\uparrow}\hat{a}_{p,\downarrow}$.
Thus, to ensure that the series of two controlled-$U$ gate operations commutes with both spin operators, we must cancel the spin-polarization contribution by enforcing equal parameterized angles for each spin-orbital, \textit{i.e.}, $\tilde \theta = \theta_\uparrow = \theta_\downarrow$. 
This simplifies the resulting unitary operation to
\begin{equation} \label{eq:exp_commuting}
    C_{p\uparrow}\hat{U}(\tilde \theta) C_{p\downarrow}\hat{U}(\tilde \theta) = \exp(\tilde \theta \hat{n}_p \otimes \hat{A}).
\end{equation}
This is precisely the choice of parametrization we adopted in all our circuit ansatz designs presented in this work, thus ensuring the preservation of the fermionic spin properties of the output states in $\hat{S}^2$ and $\hat{S}_z$.  
Interestingly, while such a demonstration can be straightforwardly applied to qubit- and qudit-based architectures, a similar reasoning can also be extended to hybrid qubit-qumode architectures with only a few additional steps.

\section{Alternative Quantum Circuits on qubit-platforms \\ for Givens and control-Givens rotations}
\label{appendix:alternative circuit}
Starting from the original definition of the Givens rotation between two qubits:  
\begin{equation}
    \hat{G}_{qq'}(\theta) = \exp\left(-i\frac{\theta}{2} (X_q \otimes Y_{q'}  - Y_q \otimes X_{q'})\right),
\end{equation}  
we can decompose this transformation using standard circuit decompositions of Pauli strings.  

First, we note that the operators satisfy the commutation relation:  
\begin{equation}
    [X_q \otimes Y_{q'} , Y_q \otimes X_{q'}] = 0,
\end{equation}  
which allows us to split the exponential into two separate Pauli-string exponentials:  
\begin{equation}
    \exp \left(-i\frac{\theta}{2}  X_q \otimes Y_{q'} \right)  
    \quad \text{and} \quad  
    \exp \left(-i\frac{\theta}{2}  Y_q \otimes X_{q'} \right).
\end{equation}  

\begin{figure}[H]
    \centering
    \includegraphics[width=\columnwidth]{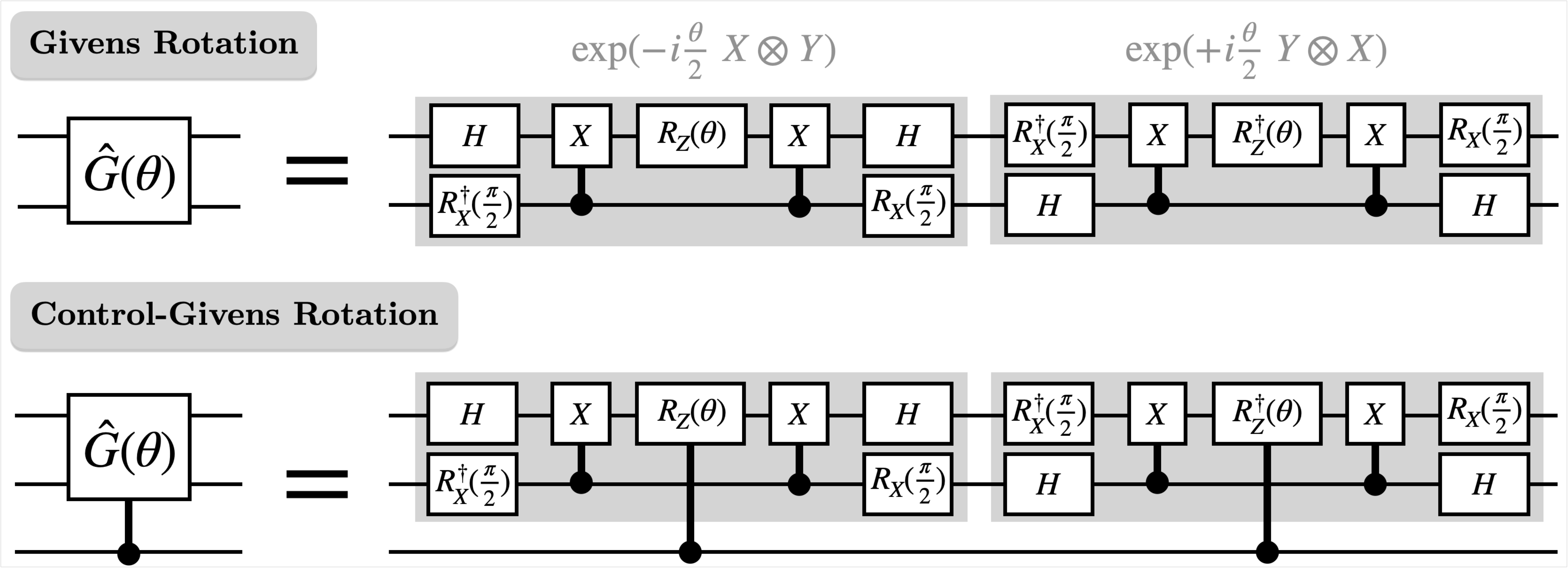}
    \caption{Illustration of alternative quantum circuits for implementing a Givens rotation (upper panel) and a controlled-Givens rotation (lower panel) based on the exponentiation of commuting Pauli-strings. The controlled-Givens rotation requires only six entangling control gates, rather than fourteen, due to circuit symmetry. This symmetry enables the cancellation of certain gates when the control qubit is in the zero state, resulting in a more compact representation.}
    \label{fig:alternative_implementation_Givens}
\end{figure}

As illustrated in Figure~\ref{fig:alternative_implementation_Givens}, we can leverage standard circuit decompositions for Pauli-string exponentiation to construct a quantum circuit for both the Givens rotation and its controlled version. While this represents one of the most conventional approaches, alternative methods may also be considered, though they generally exhibit similar gate complexity (see Ref.~\cite{google2020hartree}).

\section{ Establishing a Fair Basis for Comparing \\ Qubit, Qudit, and Qumode Platforms }
\label{app:Fair Comp}

The representation of the photonic cavity mode using different types of quantum information units inevitably leads to distinct descriptions of the physics of a polaritonic system.  
Qubits and qudits are discrete quantum variables with finite-dimensional Hilbert spaces, which naturally require truncation based on a maximum photon number $N_B^\text{max}$ when encoding the photonic mode (we typically set $N_B^\text{max} = 3$ in all our simulations).  
In contrast, qumodes employ continuous variables associated with infinite-dimensional Hilbert spaces, and can, in principle, encode the full Fock space of the cavity mode.  
While the qubit and qudit-based approach is inherently limited in representing hybrid light matter configurations due to the photon cutoff $N_B^\text{max}$, the qumode-based approach should, in principle, offer a more realistic description of the intrinsic polaritonic physics by incorporating contributions from arbitrarily high photon numbers.  
As such, drawing fair comparisons between these three platforms requires careful consideration to avoid introducing methodological bias.

To ensure a consistent and meaningful comparison across platforms, we first performed QED-FCI calculations with a large number of photons (e.g., 15 bosons) to identify the coupling regimes where the three-photon truncation used in qubit/qudit platforms remains valid. 
This analysis indicated that reliable results are confined to light–matter coupling strengths up to $\lambda = 0.25$. 
Within this range, QED-FCI shows that the three lowest polaritonic states can be accurately captured using mainly hybrid-light matter configurations containing at most $N_B^\text{max} = 3$, enabling qubit, qudit, and qumode models to produce closely matching eigenstates structure and energies. 
Going beyond this already very high value of coupling $\lambda$ (compared to what is usually investigated in the polaritonic literature), revealed important differences between the three platforms, which led to significant distortions in the polaritonic potential energy surfaces and excited state reordering.
Limiting our study to $\lambda \leq 0.25$ therefore maintains physical consistency and comparability across all three approaches considered in the present work.

\section{QED-FCI Wavefunction Analysis  \\ for Increasing Light-Matter Coupling $\lambda$}
\label{app:WF_structure}

In this appendix, we detail our numerical simulations focused on the effect of strong light-matter interaction on the internal Configuration Interaction structure of the three low lying polaritonic states as the coupling strength $\lambda$ increases.
All results presented here are obtained within the QED-FCI framework (\textit{i.e.} exact diagonalization)  assuming a maximum number of $N_B^\text{max} =3$ photons in the cavity.
To carry out our study, we considered the interatomic distance $r = 0.74 $ \AA\ for which resonance with the cavity frequency $\omega = 1$ Ha leads to the emergence of a LIAC (see Fig.~\ref{fig:H2_illustration}). 
We then systematically varied the coupling amplitude in the range $\lambda \in [0, 0.25]$.
For each coupling $\lambda$, we then evaluated the maximum (absolute) wavefunction coefficients observed in the four accessible photonic subspaces for the three low-lying eigenstates.
To construct this measure, we numerically defined specific photonic operators $\hat{P}_\alpha =   \ketbra{n_\alpha}{n_\alpha} \otimes \bold{1}_\text{elec}$ which project onto the Hilbert space conserving the photon number ($n_\alpha = 0$, 1, 2, or 3). 
We then used these operators to project each eigenstate onto the corresponding subspace and identified the maximum wavefunction amplitude over the remaining fermionic Slater determinants.

\begin{figure}[t]
    \centering
    \includegraphics[width=\columnwidth]{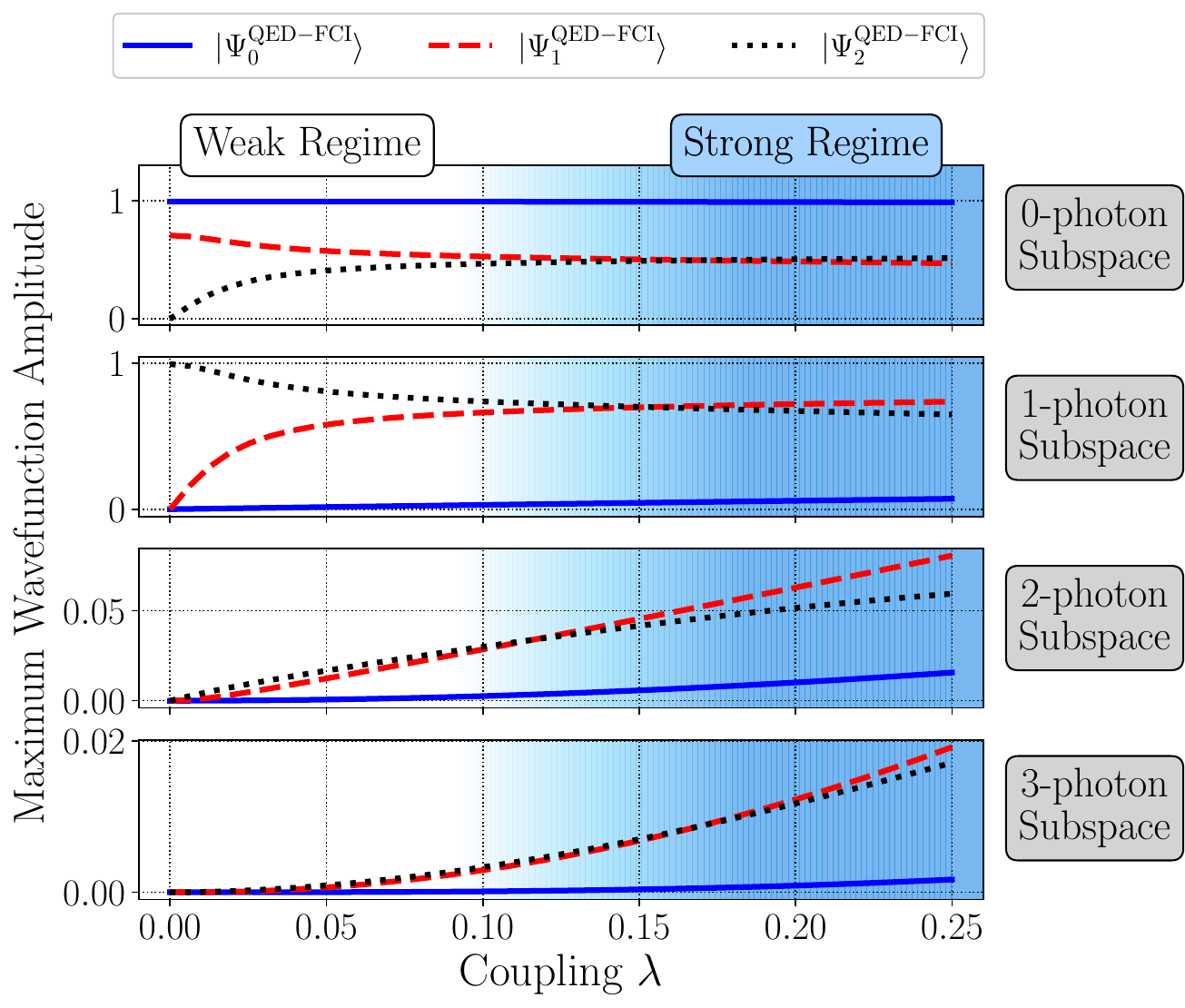} 
    \caption{\textbf{Illustration of QED-FCI eigenstates structures as a function of the light-matter coupling parameter $\lambda$.} 
    All the results presented here are obtained via exact diagonalization of the H$_2$ Pauli-Fierz Hamiltonian. 
    From top to bottom respectively represent the maximum amplitudes (in absolute value) of the three low lying eigenstates of the polaritonic system obtained in the 0-, 1-, 2- and 3-photon subspaces. 
    The interatomic distance is fixed to $r=0.74$\ \AA.
    } 
    \label{fig:WF_decomposition}
\end{figure}

In Figure~\ref{fig:WF_decomposition}  we illustrate the results of our numerical investigations.
More precisely, the top, middle, and bottom panels represent the 0-, 1-, 2- and 3-photon subspaces, respectively.
The maximum wave function amplitudes encountered for the ground state are plotted as solid blue lines, as dashed red lines for the first excited state, and as dotted black lines for the second excited state.
As clearly seen here, the structure of the exact eigenstates strongly depends on the coupling regime.
In the absence of coupling $\lambda = 0$, the exact ground state and the first excited state reside entirely in the 0-photon subspace (top panel, solid blue curve and red dashed curve), while the second excited state resides solely in the 1-photon subspace (second panel, black dotted curve).
There is no contribution from the 2- and 3-photon subspace in this regime (all wavefunction amplitudes at zero in the third and fourth panels).
However, as the coupling parameter $\lambda$ increases to reach stronger regimes, the structure of the three eigenstates changes significantly, leading to non-zero components in subspaces that were previously unoccupied.
Notably, all three eigenstates acquire non-zero contributions in the 2- and 3-photon subspaces (see third and fourth panel).
In addition, both the ground and first excited states gain amplitude in the 1-photon subspace (second panel, solid blue curve and red dashed curve), while the second excited state acquires components in the 0-photon subspace (top panel, black dotted curve).
All these eigenstate restructuring are consequences of the arising of nontrivial electron-photon correlation mechanisms generated by coupling with the cavity.


\bibliography{biblio.bib}

\end{document}